\documentclass[11pt]{article}
\usepackage[utf8]{inputenc}
\usepackage[T1]{fontenc}
\usepackage{amsmath, amssymb, amsthm, mathtools}
\usepackage{booktabs}
\usepackage{graphicx}
\usepackage{xcolor}
\usepackage{hyperref}
\usepackage{natbib}
\usepackage{enumitem}
\usepackage{algorithm}
\usepackage{algpseudocode}
\usepackage{caption}
\usepackage{tikz}
\usetikzlibrary{arrows.meta, positioning, shapes.geometric, fit, calc, decorations.pathreplacing}
\usepackage{geometry}
\newtheorem{definition}{Definition}

\newtheorem{remark}{Remark}

\title{Selective Control under Noisy Perception: Governance Failures Hidden by Aggregate Metrics in Modular Networks}
\author{Igor Itkin\\
\small Independent Researcher, Tel Aviv, Israel\\
\small \texttt{ig.itkin@gmail.com}\\
\small ORCID: \href{https://orcid.org/0009-0004-9513-8463}{0009-0004-9513-8463}}
\date{May 2026}

\begin{document}
\maketitle

\begin{abstract}
A content-moderation system can score well on every standard accuracy metric and still cause real harm, if its mistakes fall on the few users who connect otherwise separate communities. We show this in an agent-based model where $N{=}240$ learning agents on a community-structured network each post harmless, productive, or dangerous content, and a regulator removes or penalizes whatever a noisy classifier flags. Overall usefulness barely moves as the noise changes (one-way ANOVA, $p{=}0.96$): by aggregate measures, nothing looks wrong. The damage instead concentrates on these bridge users, whose useful posts are wrongly suppressed and whose dangerous posts are wrongly spared. A governance loss ($\mathcal{L}_{\mathrm{gov}}$) that prices these two mistakes separately from the cost of enforcement more than doubles under false-positive-heavy noise. Aggregate accuracy hides who is harmed, and the cheap quantity to audit is how many connections a user has (degree), a near-perfect proxy for the betweenness that defines a bridge ($r = 0.96$).
\end{abstract}

\section{Introduction}
\label{sec:introduction}

Online platforms and regulatory institutions govern populations that behave less like homogeneous crowds and more like networks: participants cluster into communities (interest groups, forums, regional or topical clusters), densely connected within and only sparsely connected across. Because no institution can directly observe the true intent behind every post or account at scale, it must decide whom to act on from the label an automatic classifier (a content filter, a spam detector, a risk score) assigns, removing, throttling, or sanctioning whatever is marked harmful. The classifier is thus the institution's only window onto the population, and an imperfect one: even at high overall accuracy it mislabels a non-trivial fraction of items, and those mistakes are not spread uniformly across the network. The few nodes that bridge otherwise separated communities carry disproportionate weight \citep{kempe2003maximizing}, because cross-community influence must pass through them. A false negative on such a bridge, a missed dangerous item, lets harmful content cross a structural barrier that would otherwise contain it, whereas a false positive, a wrongly flagged productive item, severs the coordination that links the two communities; the same error deep inside a single community is far more easily absorbed. Standard evaluation metrics (precision, recall, the F1 score, aggregate system performance) average over all nodes and so cannot see where an error lands. An audit built on them, whether it asks about fairness or about safety, certifies a governance regime as sound by every aggregate measure exactly when its harms are concentrating at the positions that hold the network together (Figure~\ref{fig:mechanism}). The quantity an audit should track is therefore where the classifier's errors fall, not how often it errs.

\begin{figure}[t]
\centering
\begin{tikzpicture}[
    node distance=0.8cm and 1.6cm,
    box/.style={draw, rounded corners, minimum width=2.0cm, minimum height=0.6cm, align=center, font=\small},
    arr/.style={-{Stealth[length=5pt]}, thick},
    label/.style={font=\footnotesize\itshape, text=gray!70!black},
]

\begin{scope}[local bounding box=left]
\node[font=\small\bfseries] (titleL) {(a) Oracle classification};
\node[box, below=0.4cm of titleL] (classL) {Classifier\\$\mathbf{M}=\mathbf{I}$};
\node[box, right=1.2cm of classL] (bridgeL) {Bridge nodes\\correctly labeled};
\node[box, below=0.6cm of classL] (aggL) {Aggregate\\usefulness};
\node[box, below=0.6cm of bridgeL] (govL) {Governance\\loss};

\draw[arr] (classL) -- node[above, label] {perfect} (bridgeL);
\draw[arr] (classL) -- (aggL);
\draw[arr] (bridgeL) -- (govL);

\node[below=0.2cm of aggL, font=\small, text=green!50!black] {$2.21$};
\node[below=0.2cm of govL, font=\small, text=green!50!black] {$0.039$};
\end{scope}

\begin{scope}[xshift=8.5cm, local bounding box=right]
\node[font=\small\bfseries] (titleR) {(b) FP-heavy noise};
\node[box, below=0.4cm of titleR] (classR) {Classifier\\64\% accurate};
\node[box, right=1.2cm of classR] (bridgeR) {Productive bridges\\mislabeled $P{\to}D$};
\node[box, below=0.6cm of classR] (aggR) {Aggregate\\usefulness};
\node[box, below=0.6cm of bridgeR] (govR) {Governance\\loss};

\draw[arr, red!70!black, dashed] (classR) -- node[above, label, text=red!70!black] {noisy} (bridgeR);
\draw[arr] (classR) -- (aggR);
\draw[arr, red!70!black] (bridgeR) -- (govR);

\node[below=0.2cm of aggR, font=\small, text=green!50!black] {$2.21$ (same)};
\node[below=0.2cm of govR, font=\small, text=red!60!black] {$0.088$ ($2.3\times$)};
\end{scope}

\draw[gray, dashed] ($(left.north east)+(0.5,0.3)$) -- ($(left.south east)+(0.5,-0.3)$);

\end{tikzpicture}
\caption{The bridge-error dilution mechanism. (a)~Under oracle classification, all labels are correct; governance loss consists only of control cost. (b)~Under false-positive-heavy noise, productive bridge nodes are mislabeled as dangerous and suppressed. Aggregate usefulness is unchanged because bridges are a small fraction of the population. \textbf{Conclusion:} position-specific errors are invisible to global metrics but detectable by a bridge-weighted governance loss.}
\label{fig:mechanism}
\end{figure}

Two research traditions speak to this problem but rarely meet. Evolutionary game dynamics on networks \citep{szabo2007evolutionary, perc2017statistical} and enforcement in agent-based models \citep{epstein2002civil, salahshour2022cost} study how punishment and structure shape collective behavior, but they assume the regulator observes the true state of each agent. Work on algorithmic content moderation \citep{gorwa2020algorithmic, gillespie2020content} takes classifier error seriously, yet evaluates it with position-blind aggregate metrics. Neither tradition asks what happens when classification error is filtered through network position, nor how that interacts with an enforcement policy that adapts \citep{zheng2022aieconomist}, a feedback signal that arrives late, or content that itself changes in response to enforcement. Section~\ref{sec:related_work} develops these threads in detail; the gap they leave is a governance evaluation sensitive to \emph{where} errors fall.

We make the problem concrete with a deliberately minimal bilevel model: a regulator acting on a population that adapts in response. Agents occupy a stochastic block model (SBM) network, a graph whose nodes split into densely linked \emph{communities} joined by a few \emph{bridge} nodes, and each agent produces content of one of three types: \emph{harmless}, \emph{productive} (valuable cross-agent coordination), or \emph{dangerous}. Agents adjust their behavior over time through tabular Q-learning. The regulator never sees these true types; it sees only noisy classifier labels, and it punishes whatever is labeled dangerous, optionally punishing flagged bridge nodes more harshly, an intensity we call its \emph{bridge targeting}. We judge the outcome through two lenses. \emph{Aggregate usefulness} is the conventional one: the total value the population produces, computed from agents' true content types. \emph{Governance loss}, $\mathcal{L}_{\mathrm{gov}}$, is the metric we propose: a position-weighted cost of the two ways governance fails---dangerous activity left unchecked at bridges, and productive activity wrongly suppressed at bridges---plus the cost of control itself. A \emph{governance failure}, throughout, means one of those two priced events. The abstraction is intentional, in the tradition of stylized social simulation \citep{epstein2002civil}: the aim is not predictive realism but an isolated, reproducible demonstration of failures that aggregate usefulness cannot see. The study is organized around five research questions:

\begin{enumerate}
\item[\textbf{RQ1.}] Can aggregate usefulness detect governance failures when classification errors concentrate at bridge nodes rather than spreading uniformly?
\item[\textbf{RQ2.}] Does the governance loss separate the distinct failure modes that aggregate usefulness conflates?
\item[\textbf{RQ3.}] Can a regulator that adapts its bridge-targeting intensity resolve the tradeoff between catching dangerous content and suppressing productive content?
\item[\textbf{RQ4.}] Do institutional delay (the lag between an agent's action and the regulator's response) and classification noise compound at bridge positions, or act independently?
\item[\textbf{RQ5.}] How does endogenous content (content types that evolve with agent behavior) change the governance picture?
\end{enumerate}

We make five contributions, one per research question:

\begin{enumerate}
\item A governance loss, $\mathcal{L}_{\mathrm{gov}} = \mathcal{L}_{\mathrm{FN}} + \mathcal{L}_{\mathrm{FP}} + \mathcal{L}_{\mathrm{control}}$, that separates three failure modes (dangerous activity missed at bridges, productive activity suppressed at bridges, and control expenditure) and is built to satisfy four stated desiderata: exposure weighting, bridge sensitivity, convex control cost, and decomposability (RQ2).
\item A demonstration that governance loss registers position-concentrated classification errors that aggregate usefulness averages away. Usefulness stays within $2.211$--$2.216$ across noise regimes (one-way analysis of variance, $p=0.96$); governance loss more than doubles under false-positive-heavy noise ($0.039 \rightarrow 0.088$, Cohen's $d=2.41$), and the ordering holds across all 27 weighting configurations we test. Whether those errors are also more \emph{costly} we test directly: with one-hop influence (Experiment~10) bridge-placed danger spreads no further than danger elsewhere, but adding a multi-hop contagion (Experiment~11) makes structural position carry large consequence (Cohen's $d > 3$) in the non-saturating regime, at inter-community density high enough for a cascade to cross, operating through node degree, for which betweenness is a near-collinear proxy in modular networks (RQ1).
\item An adaptive bridge-targeting bandit that fails to ease off under noisy classification. This exposes a reward-design pitfall: adaptive governance needs a reward aligned with governance loss, not with raw bridge error rates (RQ3).
\item A delay-by-noise study showing that institutional delay and classification noise act through independent pathways rather than compounding. Delay drives instability through the alarm feedback loop; noise drives governance failure through bridge errors; a regulator facing both must address them separately (RQ4).
\item An endogenous-content extension in which content type evolves with behavior, coupling enforcement, behavior, and the classifier's ground truth in a feedback loop that the fixed-type model cannot capture (RQ5).
\end{enumerate}

The contribution is methodological: a metric and a set of mechanisms, demonstrated in a controlled model rather than calibrated to a platform. Section~\ref{sec:related_work} reviews related work; Sections~\ref{sec:model}--\ref{sec:governance_loss} define the model and the governance loss; Section~\ref{sec:experiments} reports eleven experiments answering the research questions; Section~\ref{sec:discussion} discusses implications and limitations. The companion paper treats institutional delay without classification noise. Code and data are available at \url{https://github.com/YehudaItkin/noisy-perception-governance}.\footnote{The repository will be made public upon publication.}

\section{Related Work}
\label{sec:related_work}

\paragraph{Enforcement on networks.}
Network topology shapes cooperation dynamics \citep{szabo2007evolutionary, nowak2006five}: punishment and reward mechanisms alter equilibrium selection depending on population structure \citep{sigmund2001reward, perc2013evolutionary}, and modular community organization creates heterogeneous roles for different positions \citep{holland1983stochastic, girvan2002community}. \citet{epstein2002civil} introduced the canonical agent-based model of civil violence, where enforcement level affects rebellion nonlinearly and generates punctuated equilibrium. \citet{siegel2011repression} demonstrated that repression efficacy is nonmonotonic: the same enforcement level that suppresses collective action in one topology triggers backlash in another. More recently, \citet{salahshour2022cost} showed experimentally that as punishment noise increases, contributions decrease and punishment \emph{intensifies}, producing a 45\% drop in gains, a destructive feedback loop they call ``the cost of noise.'' \citet{zhao2024cooperation} found that Q-learning agents under probabilistic punishment exhibit both continuous and discontinuous cooperation phase transitions. The premise that bridge nodes carry outsized influence over cross-community spread is grounded in the influence-maximization literature \citep{kempe2003maximizing}, which shows that high-centrality nodes disproportionately determine how far a behavior propagates through a network. This literature establishes that enforcement outcomes depend on \emph{where} enforcement is applied and how accurately it targets, but does not combine network-position-dependent errors with noisy classification.

\paragraph{Adaptive regulators and bilevel learning.}
A regulator that adapts its targeting policy based on observed outcomes faces a bilevel learning problem: agents best-respond to the regulator while the regulator learns from agent behavior. \citet{zheng2022aieconomist} demonstrated this in the AI Economist, where a social planner and economic agents co-adapt via two-level deep reinforcement learning. \citet{yang2022adaptive} proposed meta-gradient incentive design for inducing cooperation among selfish agents, and \citet{curry2025adage} formalized bilevel ABM as a Stackelberg game in the ADAGE framework. This connects to the broader literature on inspection games and Stackelberg security games \citep{avenhaus2002inspection, tambe2011security}, where a defender allocates limited enforcement across targets that an adversary may exploit, and detection is imperfect. Our setting differs in that the regulator faces classification noise rather than strategic evasion, and the targets are network positions rather than discrete assets. At a smaller scale, \citet{henderson2023audit} framed IRS audit selection as an explore-exploit bandit where audit policy must simultaneously maximize detection reward and maintain unbiased population estimates. Our adaptive multiplier bandit (Section~\ref{sec:model}) is closest to this last line: the regulator adjusts bridge targeting intensity based on observed bridge error rates, trading off suppression of dangerous bridges against collateral damage to productive ones.

\paragraph{Content moderation and algorithmic governance.}
Algorithmic content moderation introduces position-independent classification errors into governance decisions \citep{gorwa2020algorithmic, gillespie2020content}. Evaluation of these systems typically relies on aggregate accuracy metrics that average over all nodes and content types, treating a false positive on an isolated peripheral node identically to one on a high-betweenness bridge. \citet{truong2025delayed} showed in a calibrated ABM that delayed takedown of illegal content produces steep, nonlinear degradation of moderation effectiveness: moderation becomes negligible when delay exceeds roughly two weeks. \citet{bail2018exposure} showed empirically that exposure to opposing views can increase polarization, the backfire effect. Our observation that aggressive bridge targeting trends self-defeating under noisy classification is the enforcement analog: a policy designed to suppress danger at critical positions can instead suppress productive coordination when classification is inaccurate.

\paragraph{Opinion dynamics and endogenous type change.}
\citet{flache2017opinion} survey opinion dynamics models and discuss external and institutional influence among the open directions. \citet{banisch2019polarization} showed that reinforcement-learning agents on modular networks develop stable polarization through social feedback, a mechanism structurally parallel to our Q-learning agents on stochastic block models. \citet{sobkowicz2018opinion} modeled opinion evolution via Bayesian updating under fixed cognitive biases, showing how persistent biases shape which opinions stabilize. \citet{hegselmann2002opinion} introduced the bounded-confidence model in which opinions converge through repeated averaging among sufficiently similar agents; later extensions add steering signals that bias this convergence. Our endogenous content extension (Section~\ref{sec:model}) draws on this tradition: content type transitions depend on agent behavior, creating a feedback loop between enforcement, adaptation, and the classifier's ground truth.

\paragraph{Delay and noise in evolutionary dynamics.}
The companion paper addresses delayed repression without classification noise. The interaction between delay and stochastic perturbations is theoretically subtle: \citet{miekisz2011stochastic} showed that the compound effect is model-dependent: in some update rules delay and noise compound to destabilize the ESS, while in others the slowed dynamics make the ESS robust to both. \citet{miekisz2025intrinsic} found that in structured replicator dynamics with delays interpreted as juvenile compartments, stochastic dynamics yield \emph{higher} cooperation than deterministic dynamics. \citet{dasgupta2007coordination} showed that in global games, strategic delay under noisy signals partially compensates for noise by allowing agents to gather more information. In feedback-evolving games, \citet{yan2021cooperator} showed that delayed environmental feedback combined with punishment generates oscillations through a Hopf mechanism, a parallel finding where delay acts on agent contributions rather than institutional observation. \citet{ke2025adaptive} introduced adaptive punishment that dynamically adjusts intensity to the fraction of cooperators, producing limit cycles and Hopf bifurcations reminiscent of our adaptive bandit results. Our Experiment~5 tests the delay--noise compound effect in a setting these papers do not address: noisy classification at structurally important bridge positions combined with institutional observation delay.

\paragraph{Gap.}
Governance loss is a position-weighted, cost-sensitive error metric, and as such inherits the logic of cost-sensitive learning \citep{elkan2001foundations}, which weights misclassifications by their consequences rather than counting them uniformly. It differs in two ways: the weights are derived from network structure (betweenness) rather than a fixed cost matrix, and the metric is decomposed into named failure modes (missed threats, suppressed coordination, control cost) rather than collapsed into a single expected cost. Fairness-aware metrics for graph-based classifiers address group-level disparities \citep{kang2022rawlsgcn} but do not decompose errors by structural network position or by failure mode. More broadly, none of the threads above combines network-position-dependent classification errors with adaptive enforcement, endogenous content dynamics, and institutional delay on modular networks. How these interact to produce governance failures invisible to standard metrics---and whether adaptive targeting can mitigate them---has not been studied. The governance loss metric $\mathcal{L}_{\mathrm{gov}}$ (Section~\ref{sec:governance_loss}) addresses the diagnostic side; the adaptive multiplier bandit and endogenous content extension address the prescriptive and dynamic sides.

\section{Model}
\label{sec:model}

We model a population of adaptive agents on a modular network together with a regulator that observes noisy content labels and enforces selectively. The formulation below builds up from the network and agent behavior to the enforcement rule, then adds the two optional mechanisms---adaptive bridge targeting and endogenous content---that later experiments switch on and off.

\subsection{Problem formulation}

We consider a regulator governing a population of $N$ agents on a modular network under imperfect observation. Formally:
\begin{itemize}
\item \textbf{Input:} network $G$, confusion matrix $\mathbf{M}$, bridge targeting multiplier $m$, agent architecture (Q-learning).
\item \textbf{Observation:} the regulator sees predicted content labels $\hat{c}_i$, not true types $c_i$.
\item \textbf{Action:} selective enforcement with intensity $p_i(t)$ depending on predicted label and network position.
\item \textbf{Question:} does the standard aggregate evaluation metric (usefulness) detect governance failures that arise from position-dependent classification errors?
\end{itemize}
We formalize the model components below and define the governance loss metric in Section~\ref{sec:governance_loss}.

\subsection{Network structure}

Agents occupy positions in a directed modular graph $G = (V, E)$ generated by a stochastic block model \citep{holland1983stochastic}; edges are directed to represent asymmetric influence, but betweenness centrality is computed on the undirected projection. The network consists of $n$ agents partitioned into $K$ communities with intra-community edge probability $p_{\mathrm{in}}$ and inter-community edge probability $p_{\mathrm{out}} \ll p_{\mathrm{in}}$. This construction produces modular community structure of the kind studied by \citet{girvan2002community}, where dense internal connectivity coexists with sparse inter-community links.

Bridge nodes $\mathcal{B} \subset V$ are defined as the set of nodes whose betweenness centrality \citep{freeman1977betweenness} exceeds a threshold percentile within the network. Formally, the betweenness centrality of node $v$ is $g(v) = \sum_{s \neq v \neq t} n_{st}(v) / n_{st}$, where $n_{st}$ is the total number of shortest paths from $s$ to $t$ and $n_{st}(v)$ is the number of those paths passing through $v$. Nodes in the top quantile of betweenness centrality are designated as bridges.\footnote{For Erd\H{o}s--R\'{e}nyi and scale-free topologies used in the cross-topology experiment (Experiment~2), bridge nodes are identified by degree centrality rather than betweenness centrality, since degree correlates strongly with betweenness in these non-modular topologies and is computationally cheaper.} These nodes disproportionately mediate information flow between communities and therefore play an outsized role in both productive coordination and dangerous contagion.

\subsection{Agent behavior and content types}

Each agent $i$ selects an action $a_i(t) \in \{L, M, R\}$ at each discrete time step, corresponding to loyal (safe), moderate, and radical behavior respectively. Actions yield immediate payoffs that increase with risk: the benefit of action $R$ exceeds that of $M$, which exceeds that of $L$. Agents adapt their action selection through tabular Q-learning \citep{sutton2018reinforcement} with $\epsilon$-greedy exploration, maximizing expected discounted reward $\mathbb{E}\sum_{t}\gamma^t r_i(t)$, where the per-step reward $r_i(t) = b(a_i(t), c_i) - \mathbb{1}[\text{punished}]\,\kappa(a_i(t))$ combines an action/content benefit $b$ with the punishment cost $\kappa$ incurred when the regulator acts on agent $i$ (the same agent reward as the companion paper; benefit and cost values are listed in Appendix~\ref{sec:appendix_details}). The state space encodes the local alarm level observed by the agent.

Each agent also carries a fixed content type $c_i \in \{H, P, D\}$, representing harmless content, productive autonomy, and dangerous radicalization respectively. Content type determines the true nature of an agent's output but is not directly observable by the regulator. The distinction between action (behavioral choice) and content type (latent category) is fundamental: an agent choosing radical action $R$ may be producing productive innovation ($c_i = P$) or genuinely dangerous content ($c_i = D$), and the regulator must classify this distinction under uncertainty.

\subsection{Noisy classification}

The regulator does not observe content types directly. Instead, it receives a noisy predicted label $\hat{c}_i(t)$ drawn from a confusion matrix $\mathbf{M}$ where entry $M_{xy} = \Pr(\hat{c} = y \mid c = x)$. The confusion matrix is sampled once per node per time step and cached, so that all downstream decisions (observation, enforcement, and metric computation) reference the same noisy realization.

Four noise regimes are studied. The oracle regime sets $\mathbf{M} = \mathbf{I}$, providing perfect classification. The default noise regime introduces moderate off-diagonal entries reflecting asymmetric classification uncertainty. The FP-heavy regime inflates the misclassification rate for both harmless ($H{\to}D$: $0.02{\to}0.20$) and productive ($P{\to}D$: $0.15{\to}0.35$) content, which raises false positives across non-dangerous nodes. The FN-heavy regime inflates the probability that dangerous content ($D$) is misclassified as harmless ($H$) or productive ($P$), which raises false negatives on dangerous nodes.

\subsection{Selective enforcement}

The regulator applies punishment to agents based on their predicted labels and network position. The punishment probability for agent $i$ at time $t$ takes the form
\begin{equation}
\label{eq:punishment}
p_i(t) = u_t \cdot \sigma\bigl(k(A(t - \Delta) - A_c)\bigr) \cdot d_i \cdot m_i(\hat{c}_i),
\end{equation}
where $u_t$ is the control intensity, $\sigma(\cdot)$ is a sigmoid activation, $A(t - \Delta)$ is the alarm level observed with institutional delay $\Delta$, $A_c$ is the alarm threshold, $d_i$ is a detectability factor, and $m_i(\hat{c}_i)$ is a bridge targeting multiplier that can be elevated for bridge nodes perceived as dangerous.

The bridge targeting multiplier implements selective enforcement: when $m_i > 1$ for bridge nodes classified as dangerous, the regulator applies disproportionate scrutiny to structurally central positions. The intent of the policy is to reduce cascade risk when classification is accurate, at the price of collateral costs when false positives are frequent; whether this tradeoff materializes is an empirical question (Section~\ref{sec:experiments}). The multiplier is swept across values $\{1.0, 1.35, 1.8\}$ in experiments, where $m_i = 1.0$ corresponds to uniform enforcement and $m_i = 1.8$ corresponds to aggressive bridge targeting.

\subsection{Adaptive bridge targeting}
\label{sec:adaptive_mult}

The static bridge multiplier $m$ faces a potential dilemma: the policy that works best under accurate classification may do the most damage under noisy classification. In our experiments this dilemma appears as a directional trend rather than a significant effect (Section~\ref{sec:experiments}), but it motivates the question of adaptation: an adaptive regulator should increase $m$ when dangerous content concentrates at bridge positions and decrease $m$ when false positives dominate.

We implement this as an epsilon-greedy bandit over six discrete multiplier levels, $m$ from $1.0$ to $2.0$ in steps of $0.2$. The bandit maintains an exponential moving average (EMA) of reward per level with smoothing parameter $\alpha_{\mathrm{EMA}} = 0.05$. At each step, it selects the level with the highest EMA reward with probability $1 - \epsilon$ and explores uniformly with probability $\epsilon = 0.10$. The reward signal penalizes both bridge false positives and undetected dangerous bridges:
\begin{equation}
\label{eq:mult_reward}
r_m(t) = -\mathrm{FP}_{\mathcal{B}}(t) - 2 D_{\mathcal{B}}(t),
\end{equation}
where the asymmetric weight ($2\times$ on missed threats) biases the bandit toward higher $m$ when danger is real; the $-\mathrm{FP}_{\mathcal{B}}$ term is meant to make it back off when false positives dominate, although Experiment~4 shows this intention fails, because $\mathrm{FP}_{\mathcal{B}}$ is set by the confusion matrix and does not respond to $m$. The bandit operates independently of the force-level Q-learning (when active), avoiding the combinatorial explosion of a joint action space.

When `adaptive\_multiplier' is disabled, the multiplier is fixed from configuration, recovering the static model of Experiments~1--3.

\subsection{Endogenous content dynamics}
\label{sec:endogenous_content}

In the base model, each agent's content type $c_i \in \{H, P, D\}$ is fixed at initialization. This is a simplifying assumption: in practice, an agent choosing radical behavior may shift toward genuinely dangerous content over time. We relax this by introducing action-contingent content transitions.

At each step, after an agent selects action $a_i(t)$, its content type transitions according to a Markov chain $\mathbf{T}^{(a)}$ where $T^{(a)}_{xy} = \Pr(c_{t+1} = y \mid c_t = x, a_t = a)$. The key rates: radical action ($R$) increases the probability of transitioning to dangerous content ($P \to D$: 0.15), while loyal action ($L$) increases the probability of transitioning to harmless content ($D \to H$: 0.08). Moderate action ($M$) produces mild drift toward productive content. All transition matrices are row-stochastic and can be scaled by a strength parameter $\beta$ that controls the overall transition speed.

This creates a feedback loop absent in the base model: enforcement based on noisy classification changes agent behavior via Q-learning, which changes content types via transitions, which changes the classification problem for the next step. Under FP-heavy noise, the loop may produce a false-alarm spiral: productive bridges are suppressed, shift to loyal behavior, and transition toward harmless content. This reduces actual danger while governance loss remains high, because the regulator continues to suppress based on stale classification patterns.

When endogenous content is disabled, content types are fixed at initialization, recovering the base model.

\subsection{Dangerous-content contagion}
\label{sec:cascade}

The governance loss weights errors by bridge position on the premise that bridges govern cross-community spread. The base model contains no process that could make this premise true. Influence in the base model is one-hop: an agent's reward depends only on its immediate neighbors (Equation~\ref{eq:punishment} and the influence term below it). A one-hop quantity cannot depend on betweenness, which is a multi-hop, global property of a node's position on shortest paths. For structural position to carry dynamic consequence, the model needs a process that propagates over more than one edge. We add one as an optional mechanism, switched on only in Experiment~11.

When the cascade is enabled, dangerous content spreads along edges by a threshold rule. A susceptible node $i$ (one whose content type is not $D$ and that has not been immunized) adopts $D$ with probability $\beta_c$ if at least a fraction $\theta$ of its in-neighbors already carry $D$:
\begin{equation}
\label{eq:cascade}
\Pr(c_i \to D) = \beta_c \cdot \mathbb{1}\!\left[\frac{|\{j \in \mathrm{pred}(i): c_j = D\}|}{|\mathrm{pred}(i)|} \ge \theta\right].
\end{equation}
The threshold $\theta$ spans the simple-to-complex contagion axis of \citet{centola2007complex}: at $\theta = 0$ a single dangerous neighbor suffices, so the contagion crosses individual bridge ties (simple contagion); at larger $\theta$ a node needs a critical mass of dangerous neighbors, which a lone cross-community tie cannot supply (complex contagion). Enforcement provides containment: a node that is both detected (predicted $D$) and punished in a step is immunized and its content reset, so missed dangerous nodes (false negatives) are the ones left free to spread. This couples the classifier's bridge-level errors to a genuine multi-hop process, the only channel in the model through which a node's structural position can affect population-level outcomes.

When the cascade is disabled, content spreads through no edges and the base model is recovered.

\subsection{Simulation loop}

Algorithm~\ref{alg:p2_sim} summarizes the complete simulation loop and shows how noisy classification feeds into selective enforcement.

\begin{figure}[ht]
\centering
\begin{minipage}{0.92\linewidth}
\begin{algorithmic}[1]
\small
\State \textbf{Input:} graph $G$, confusion matrix $\mathbf{M}$, multiplier $m$, horizon $T$, seeds
\State Initialize Q-tables $Q_i(s,a) \leftarrow 0$ for all agents $i$
\For{$t = 0, 1, \ldots, T-1$}
    \For{each agent $i$}
        \State Draw predicted label: $\hat{c}_i(t) \sim \mathbf{M}[c_i, \cdot]$ \hfill\emph{(noisy classification)}
        \State Select action $a_i \leftarrow \epsilon\text{-greedy}(Q_i, s_i, \epsilon)$
    \EndFor
    \State Compute alarm: $A(t) \leftarrow \frac{1}{N}\sum_i d_i \cdot f(a_i)$
    \State Compute enforcement: $p_i(t) \leftarrow u_t \cdot \sigma(k(A(t{-}\Delta) - A_c)) \cdot d_i \cdot m_i(\hat{c}_i)$
    \State Sample punishment: $\xi_i(t) \sim \mathrm{Bernoulli}(p_i(t))$
    \State Compute rewards and update Q-tables
\EndFor
\State \textbf{Output:} time series, governance loss $\mathcal{L}_{\mathrm{gov}}$, regime label
\end{algorithmic}
\end{minipage}
\captionof{algorithm}{Simulation loop for one run. The key addition over the companion paper's loop is the noisy classification step (line~5): each agent's true content type $c_i$ is mapped through the confusion matrix $\mathbf{M}$ to a predicted label $\hat{c}_i$, which then determines the bridge targeting multiplier $m_i(\hat{c}_i)$ in the enforcement step (line~9).}
\label{alg:p2_sim}
\end{figure}

\subsection{System metrics}

Aggregate usefulness is the mean payoff across all agents. Bridge risk is the mean radical fraction among bridge nodes specifically. The false-positive fraction on bridges, $\mathrm{FP}_{\mathcal{B}}$, is the number of non-dangerous bridge nodes incorrectly classified as dangerous divided by the total number of bridge nodes. The false-negative fraction on bridges, $\mathrm{FN}_{\mathcal{B}}$, is the number of dangerous bridge nodes incorrectly classified as non-dangerous divided by the total number of bridge nodes. These bridge-specific error rates can diverge substantially from population-level averages because bridges constitute a small fraction of all nodes, concentrating the noise effect at structurally critical positions. Bridge nodes also receive elevated charisma ($+0.2$) and detectability ($+0.15$) relative to non-bridge nodes, on the assumption that structurally central positions carry both greater influence and greater visibility to the regulator.

\section{Governance Loss}
\label{sec:governance_loss}

Aggregate accuracy hides where errors fall. This section defines a governance loss that weights classification errors by network position and splits them into named failure modes, after first stating the properties such a metric should satisfy.

\subsection{Motivation and desiderata}

Standard evaluation of classification systems reports aggregate metrics (overall accuracy, precision, recall) that treat all errors as equally costly regardless of where they occur in a network. In modular systems with heterogeneous node roles, this aggregation obscures critical governance failures. A false negative on an isolated peripheral node has minimal system-level consequences, whereas the same error on a bridge node connecting two communities enables dangerous content to propagate across an otherwise effective structural barrier. Similarly, a false positive on a bridge node carrying productive coordination severs inter-community information flow, imposing costs disproportionate to the error's contribution to aggregate false-positive statistics.

We want a metric that makes this visible. The decomposition below quantifies three distinct failure modes, each weighted by structural importance. We require any governance quality metric $\mathcal{L}$ to satisfy four properties:

\begin{enumerate}
\item \textbf{Exposure weighting.} The cost of a classification error should scale with the activity it fails to address. When no dangerous content reaches bridge nodes, the FN component should vanish regardless of the classifier's false-negative rate; when no enforcement occurs, the FP component should vanish. This is achieved by multiplying each error rate by a measure of the relevant activity level, so that the product is zero whenever either factor is zero.
\item \textbf{Bridge sensitivity.} Errors at structurally central nodes (high betweenness centrality) must be separable from population-level errors. The metric should allow inspection of bridge-specific failure independent of the aggregate error rate.
\item \textbf{Control cost convexity.} Higher control intensity should have increasing marginal cost, to reflect diminishing returns and increasing friction from governance expenditure.
\item \textbf{Decomposability.} The false-negative, false-positive, and control-cost contributions must be separately inspectable, so that an analyst can identify which failure mode dominates.
\end{enumerate}

The product form in Equation~\ref{eq:governance_loss} is a parsimonious additive-separable structure satisfying all four desiderata. Property~1 rules out terms that depend on error rates alone (they would be nonzero even absent relevant activity). Property~2 mandates bridge-restricted rather than population-level error rates. Property~3 motivates the quadratic $\bar{u}^2$ over a linear term. Property~4 is satisfied by construction since $\mathcal{L}_{\mathrm{gov}}$ decomposes into three named additive components. We do not claim uniqueness: other product-form or risk-weighted decompositions satisfying these properties exist. We claim that this is a parsimonious representative with interpretable components.

\subsection{Formal definition}

Let $\mathcal{B} \subset V$ denote the set of bridge nodes identified by betweenness centrality \citep{freeman1977betweenness}. Define the following quantities: $\mathrm{FN}_{\mathcal{B}}$ is the false-negative fraction among bridge nodes (the number of dangerous bridge nodes misclassified as non-dangerous, divided by the total number of bridge nodes; note that the maximum value equals the fraction of dangerous nodes among bridges, approximately 0.20 in our experiments), $\mathrm{FP}_{\mathcal{B}}$ is the false-positive fraction among bridge nodes (the number of non-dangerous bridge nodes misclassified as dangerous, divided by the total number of bridge nodes; note that the maximum value equals the fraction of non-dangerous nodes among bridges, approximately 0.80 in our experiments), $D_{\mathcal{B}}$ is the dangerous-radical bridge fraction (the fraction of bridge nodes simultaneously radical and carrying dangerous content), $P^{\mathrm{pun}}$ is the overall fraction of nodes punished by the regulator, and $\bar{u}$ is the mean repression probability across all agents (i.e., the population average of $p_i(t)$ from Equation~\ref{eq:punishment}, not the regulator force $u_t$ alone).

\begin{definition}[Governance Loss]
\label{def:governance_loss}
The governance loss is
\begin{equation}
\label{eq:governance_loss}
\mathcal{L}_{\mathrm{gov}} = \underbrace{\lambda_{\mathrm{FN}} \cdot \mathrm{FN}_{\mathcal{B}} \cdot D_{\mathcal{B}}}_{\mathcal{L}_{\mathrm{FN}}} + \underbrace{\lambda_{\mathrm{FP}} \cdot \mathrm{FP}_{\mathcal{B}} \cdot P^{\mathrm{pun}}}_{\mathcal{L}_{\mathrm{FP}}} + \underbrace{\lambda_{u} \cdot \bar{u}^2}_{\mathcal{L}_{\mathrm{control}}},
\end{equation}
where $\lambda_{\mathrm{FN}}$, $\lambda_{\mathrm{FP}}$, and $\lambda_u$ are weighting coefficients. The product form ensures that each error-rate term is modulated by the actual activity it fails to address: a high false-negative rate costs little if there is negligible dangerous bridge activity to miss, and a high false-positive rate costs little if few nodes are actually punished.
\end{definition}

\subsection{Interpretation and diagnostic comparison}

The first term, $\mathcal{L}_{\mathrm{FN}}$, captures the cost of undetected dangerous activity at bridge positions. When the classifier misses dangerous content on a bridge node, the regulator fails to intervene at precisely the location where intervention would be most effective at preventing cross-community spread. This term increases when false-negative rates are elevated specifically on nodes with high betweenness centrality, regardless of the population-level false-negative rate.

The second term, $\mathcal{L}_{\mathrm{FP}}$, captures the cost of suppressing productive bridge activity. When the classifier incorrectly labels productive content as dangerous on a bridge node, enforcement action removes a node that was providing valuable inter-community coordination. The cost is structural: productive bridges carry information between communities \citep{girvan2002community}, and their suppression fragments the network.

The third term, $\mathcal{L}_{\mathrm{control}}$, captures the cost of control itself. The quadratic form $\bar{u}^2$ penalizes high control intensity regardless of targeting accuracy. This reflects the general principle that governance expenditure has opportunity costs and that excessive enforcement imposes friction on all network participants.

\begin{remark}[Structural channel for an accuracy--targeting interaction]
\label{prop:complementarity}
The product form suggests a channel through which classifier accuracy $\alpha$ and targeting intensity $m$ could interact. Parameterize the confusion matrix via $\mathbf{M}(\alpha) = \alpha \mathbf{I} + (1{-}\alpha)\mathbf{M}_0$, so $\mathrm{FP}_{\mathcal{B}}(\alpha)$ decreases in $\alpha$. The relevant cross-partial of the false-positive term is
\[
\frac{\partial^2 \mathcal{L}_{\mathrm{FP}}}{\partial \alpha \, \partial m}
= \lambda_{\mathrm{FP}} \cdot \frac{\partial \mathrm{FP}_{\mathcal{B}}}{\partial \alpha} \cdot \frac{\partial P^{\mathrm{pun}}}{\partial m},
\]
which is negative (a complementarity) only to the extent that targeting actually raises the punished fraction ($\partial P^{\mathrm{pun}}/\partial m > 0$). In our model this channel turns out to be weak: the punished fraction is nearly flat in $m$ (Spearman $\rho = 0.03$, $p = 0.38$), so the FP cross-partial is small, and it is partly offset by an opposing $\mathcal{L}_{\mathrm{FN}}$ term. Whether the net interaction is complementary is therefore an empirical question, not a structural guarantee; Experiment~9 tests it directly and finds the accuracy main effect strong but the $\alpha \times m$ interaction not statistically significant. We thus report a reliable main effect of classification accuracy rather than a complementarity result.
\end{remark}

As a diagnostic example, we compare governance loss against aggregate usefulness under the experimental conditions of Section~\ref{sec:experiments}. When $\lambda_{\mathrm{FN}} = \lambda_{\mathrm{FP}} = \lambda_u = 1$, the oracle baseline achieves $\mathcal{L}_{\mathrm{gov}} = 0.039$, consisting entirely of control cost (since $\mathrm{FN}_{\mathcal{B}} = \mathrm{FP}_{\mathcal{B}} = 0$ under perfect classification). Under FP-heavy noise, $\mathcal{L}_{\mathrm{gov}}$ rises to $0.088$ ($+0.050$ absolute, $2.3\times$ relative; Cohen's $d = 2.41$) while aggregate usefulness remains unchanged at ${\approx}2.21$. This divergence illustrates, within this model, that bridge-specific, exposure-weighted error decomposition captures structural governance information that a scalar usefulness metric averages away. We do not claim that usefulness is the only alternative; bridge-specific FP/FN rates, punishment Gini coefficients, or cross-community edge activity may also detect aspects of this effect. The contribution of $\mathcal{L}_{\mathrm{gov}}$ is its compact decomposability (Desideratum~4): it separates failure modes so that an analyst can identify whether the dominant problem is missed threats, suppressed coordination, or excessive control. The metric tells you what is going wrong, not just that something is.

\section{Experiments}
\label{sec:experiments}

The model (Section~\ref{sec:model}) and governance loss metric (Section~\ref{sec:governance_loss}) set up the central question: does noisy classification at structurally important positions produce governance failures invisible to standard evaluation? Eleven experiments (6{,}720 runs total) test this from different angles. Experiments~1--3 establish two base findings: noise hides governance failures, and $\mathcal{L}_{\mathrm{gov}}$ separates failure modes; they also surface the bridge-targeting dilemma as a directional hypothesis, tested directly in Experiments~4 and~9. Experiments~4--6 ask whether adaptive targeting, institutional delay, and endogenous content dynamics change the picture. Experiments~7--10 close open questions: coupling threshold, delay invariance, joint optimization, and whether bridge position amplifies spread. Experiment~11 returns to that last question with an explicit contagion mechanism, asking whether structural position carries dynamic consequence once dangerous content can propagate, and whether the operative property is betweenness or node degree. The experiments refine the five research questions of Section~\ref{sec:introduction} into eleven operational questions:

\begin{enumerate}
\item Does aggregate usefulness detect governance failures caused by position-dependent classification errors?
\item Does bridge targeting interact with topology and classification accuracy?
\item Does the governance loss decomposition provide information beyond simpler alternatives?
\item Can an adaptive regulator handle the targeting--accuracy tradeoff that defeats static policies?
\item Do institutional delay and classification noise compound at bridge positions?
\item Does endogenous content evolution change governance dynamics relative to fixed types?
\item At what coupling strength does aggregate usefulness become sensitive to noise regime?
\item Is the adaptive targeting bandit delay-invariant?
\item How does the optimal bridge targeting intensity depend on classifier accuracy, and are the two investments complements or substitutes?
\item Does dangerous content on bridge nodes actually amplify population-level spread, as the metric's bridge weighting assumes?
\item When dangerous content can propagate over multiple hops, does structural position carry outsized consequence, and is the operative property betweenness or node degree?
\end{enumerate}

Table~\ref{tab:p2_experiment_map} maps each experiment to its operational question; figure and table captions cite these question numbers. Relative to the introduction's research questions, questions~1, 7, 10, and~11 elaborate RQ1; question~3 is RQ2; questions~2, 4, and~9 probe RQ3; questions~5 and~8 are RQ4; and question~6 is RQ5.

\begin{table}[ht]
\centering
\small
\begin{tabular}{llll}
\toprule
Experiment & Question & Varied factor & Controlled factors \\
\midrule
Exp.~1: Noise sweep & Q1 & Noise regime & Topology, $m{=}1.0$ \\
Exp.~2: Bridge targeting & Q2 & Topology $\times$ $m$ & Noise (default) \\
Exp.~3: Loss decomposition & Q3 & Noise regime $\times$ $m$ & $\lambda{=}1$, topology \\
Exp.~4: Adaptive multiplier & Q4 & Adaptive vs.\ static $\times$ noise & Topology \\
Exp.~5: Delay $\times$ noise & Q5 & Delay $\times$ noise regime & $m{=}1.35$ \\
Exp.~6: Endogenous content & Q6 & Exo/endo $\times$ noise & $m{=}1.35$ \\
Exp.~7: Coupling threshold & Q7 & Cost multiplier $\times$ noise & Topology \\
Exp.~8: Adaptive under delay & Q8 & Delay $\times$ noise (adaptive) & Topology \\
Exp.~9: Joint optimization & Q9 & $\alpha \times m$ & Topology, delay \\
Exp.~10: Bridge contagion & Q10 & Content placement & Noise (FN-heavy) \\
Exp.~11: Cascade consequence & Q11 & Placement $\times$ threshold $\theta$ & Noise (FN-heavy), $m{=}1.0$ \\
\bottomrule
\end{tabular}
\caption{Mapping of experiments to research questions. Experiments~1--3 are the base model; 4--6 extend it; 7--11 close open questions.}
\label{tab:p2_experiment_map}
\end{table}

\subsection{Experimental setup}

\paragraph{Shared settings.} All experiments use $N=240$ agents on a modular stochastic block model graph (6 communities, intra-community edge probability $p_{\mathrm{in}}=0.08$, inter-community $p_{\mathrm{out}}=0.004$, bridge fraction 12\% by betweenness centrality). The population size $N{=}240$ is large enough for stable bridge-specific statistics (approximately 28 bridge nodes at 12\%) but small enough for exhaustive Q-table exploration within 500 steps. The 6-community modular structure produces well-separated communities with a modularity ratio of $p_{\mathrm{in}}/p_{\mathrm{out}} = 20$, which gives clearly defined bridge roles. Agent Q-learning parameters ($\alpha=0.10$, $\epsilon=0.08$, $\gamma=0.95$) follow standard practice \citep{sutton2018reinforcement}. Content types ($H$, $P$, $D$) are distributed uniformly across agents to isolate the effect of classification noise from spatial content clustering. Simulation horizon: 500 steps, 50 seeds per condition, paired by seed across noise regimes to ensure that differences reflect noise regime rather than graph realization. Governance loss uses equal weighting $\lambda_{\mathrm{FN}} = \lambda_{\mathrm{FP}} = \lambda_u = 1$ throughout; a sensitivity sweep over 27 $\lambda$ configurations is reported in Appendix~\ref{sec:robustness}.

\paragraph{Baselines.} The oracle condition ($\mathbf{M} = \mathbf{I}$, perfect classification) is the baseline for all experiments. It provides the lower bound on achievable governance loss (equal to pure control cost since $\mathrm{FP}_{\mathcal{B}} = \mathrm{FN}_{\mathcal{B}} = 0$). All effect sizes (Cohen's $d$) are computed relative to this baseline. This is a mechanism study: we compare noise regimes against perfect observation, not competing governance models from the literature.

\paragraph{Statistical reporting.} The eleven experiments run many hypothesis tests, and we apply no family-wise correction. None is needed for the primary claims: they rest either on very large effects (Cohen's $d > 2$, ANOVA $F > 40$) that survive any correction, or on nulls that we back with equivalence tests (two one-sided tests, TOST) rather than with a non-significant $p$ alone, since a non-significant difference can also be a power failure. Equivalence tests accompany the usefulness nulls (Experiments~1, 6, and~7), the placement nulls (Experiments~10 and~11), and the bandit's delay-invariance (Experiment~8). The one primary null stated without an equivalence test is the delay~$\times$~noise interaction of Experiment~5, where no single equivalence bound is natural; there we report the non-significant interaction term together with the per-condition trajectories it rests on. The within-noise directional results (Experiments~2, 4, and~9) are reported as hypotheses the data is consistent with, not as confirmed effects, and are flagged as such at each occurrence.

\subsection{Experiment~1: Noise sweep}

The first experiment compares four noise regimes (oracle, default, FP-heavy, and FN-heavy) on a fixed modular network topology. The oracle condition provides perfect classification ($\mathbf{M} = \mathbf{I}$) and serves as the reference against which governance costs are measured. The default condition applies moderate asymmetric noise with off-diagonal confusion probabilities. The FP-heavy condition inflates the misclassification rate from productive ($P$) to dangerous ($D$); this elevates false positives specifically on productive nodes. The FN-heavy condition inflates the misclassification rate from dangerous ($D$) to harmless ($H$) or productive ($P$), which elevates false negatives on dangerous nodes.

For each condition, the simulation records aggregate usefulness, bridge-specific radical fraction, population-level and bridge-specific false-positive and false-negative rates, and the full governance loss decomposition. The primary hypothesis is that aggregate usefulness will remain invariant across conditions while bridge-specific error rates and governance loss will diverge. Such a finding would confirm that standard metrics are blind to governance failures hiding in plain sight.

The central empirical finding is that aggregate usefulness is invariant across all four noise conditions (Table~\ref{tab:p2_main_results}). The oracle condition yields mean usefulness of $2.213$, while the default, FP-heavy, and FN-heavy conditions yield $2.211$--$2.216$. A one-way ANOVA confirms that these differences are not statistically significant ($F=0.10$, $p=0.96$). The maximum spread across conditions is $0.005$, or $0.2\%$ of the mean. A TOST equivalence test confirms statistical equivalence to the oracle for all conditions at a bound of $\varepsilon=0.1$ ($p_{\mathrm{TOST}} < 10^{-5}$); this bound is roughly two cross-seed standard deviations ($\mathrm{SD} \approx 0.05$), so it should be read as ruling out differences larger than the run-to-run noise floor, not as a substantively calibrated threshold. A regulator who monitors only aggregate usefulness would see nothing wrong: identical performance across noise regimes that differ dramatically in governance quality.

\begin{table}[t]
\centering
\begin{tabular}{lccccc}
\toprule
Condition & Usefulness & $\mathrm{FP}_{\mathcal{B}}$ & $\mathrm{FN}_{\mathcal{B}}$ & $\mathcal{L}_{\mathrm{gov}}$ & Cohen's $d$ vs.\ oracle \\
\midrule
Oracle        & 2.213 & 0.000 & 0.000 & 0.039 & --- \\
Default noise & 2.212 & 0.077 & 0.050 & 0.056 & 1.08 \\
FP-heavy      & 2.216 & 0.228 & 0.051 & 0.088 & 2.41 \\
FN-heavy      & 2.211 & 0.053 & 0.152 & 0.057 & 0.96 \\
\bottomrule
\end{tabular}
\caption{Governance outcomes by noise regime (50 seeds per condition). Usefulness is invariant across conditions, while bridge-specific false-positive and false-negative rates diverge by factors of three or more. \textbf{Conclusion:} governance loss $\mathcal{L}_{\mathrm{gov}}$ reveals large-effect differences (Cohen's $d$ up to $2.4$) invisible to the usefulness metric (question~1). Standard errors for $\mathrm{FP}_{\mathcal{B}}$ and $\mathrm{FN}_{\mathcal{B}}$ are below $0.005$ in all conditions.}
\label{tab:p2_main_results}
\end{table}

While usefulness remains flat, bridge-specific error rates diverge sharply. The FP-heavy condition produces $\mathrm{FP}_{\mathcal{B}} = 0.228$, a three-fold increase over the default condition ($0.077$) and elevated from the oracle baseline of zero. The FN-heavy condition produces $\mathrm{FN}_{\mathcal{B}} = 0.152$, a three-fold increase over the default ($0.050$). These divergences occur entirely within bridge-specific metrics and are invisible to any evaluation framework that aggregates across all nodes or all content types.

\begin{figure}[t]
\centering
\includegraphics[width=\textwidth]{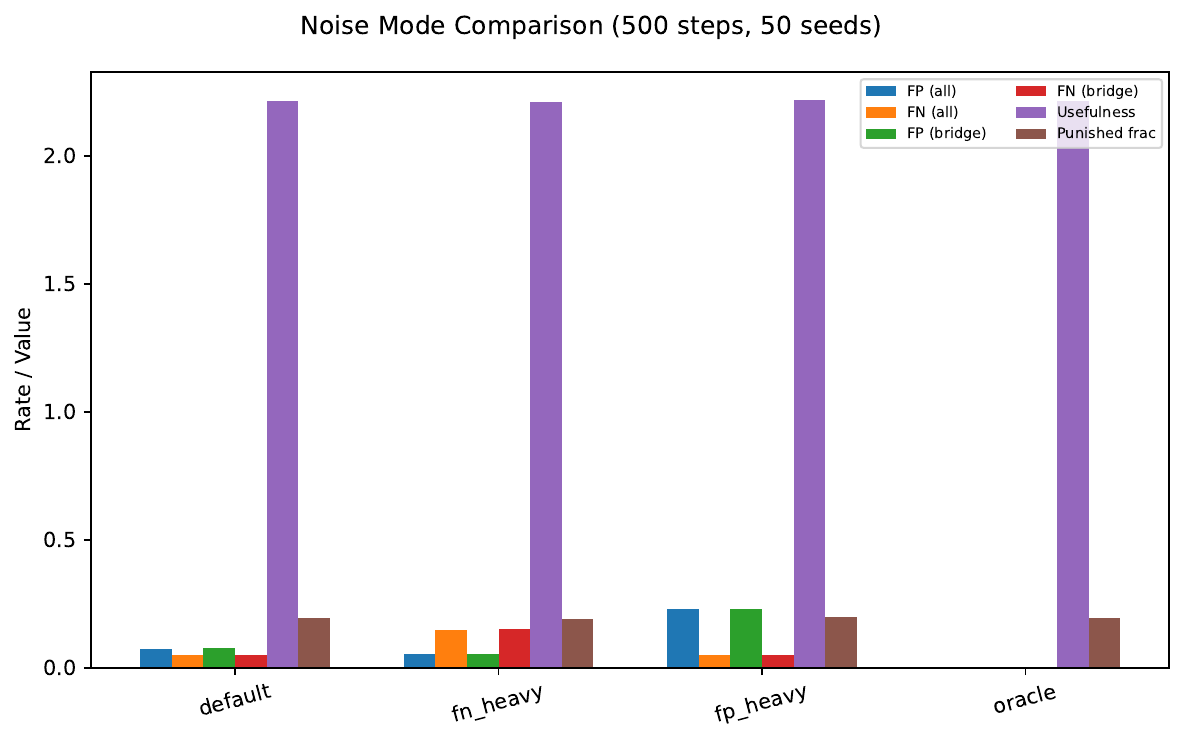}
\caption{Noise mode comparison across bridge-specific and population-level error metrics. Bridge-specific error rates diverge from population averages. \textbf{Conclusion:} governance quality concentrates at structurally central positions invisible to aggregate evaluation (question~1). Fifty seeds per condition.}
\label{fig:p2_noise}
\end{figure}

Figure~\ref{fig:p2_noise} presents the full noise sweep results: FP-heavy regimes sharply increase false-positive rates at bridge nodes, while FN-heavy regimes elevate false negatives at bridge nodes. Population-level error rates shift less dramatically than bridge-specific rates because bridge nodes represent a small fraction of the total population, and errors at bridge positions are diluted when averaged over all nodes. This dilution effect is precisely what makes bridge-specific metrics necessary for governance evaluation.

\subsection{Experiment~2: Topology and bridge targeting}

The second experiment crosses network topology (Erd\H{o}s--R\'{e}nyi, scale-free, and modular stochastic block model) with bridge targeting intensity (multiplier values $m \in \{1.0, 1.35, 1.8\}$). The uniform multiplier $m = 1.0$ applies identical enforcement pressure to all nodes regardless of structural position. Elevated multipliers concentrate enforcement on bridge nodes classified as dangerous, implementing selective governance.

This experiment tests the hypothesis that bridge targeting reduces bridge radicalization in modular networks where community structure creates clearly defined bridge roles, but that the benefit comes at the cost of increased collateral punishment, particularly when false positives cause productive bridges to be misidentified as dangerous. The modular topology is expected to show the strongest bridge effects because its community structure produces nodes with high betweenness centrality that occupy clearly differentiated structural positions. One caveat applies to the comparison: bridges are operationalized by betweenness centrality in the modular model but by degree centrality in the Erd\H{o}s--R\'{e}nyi and scale-free graphs (Section~\ref{sec:model}), so the contrast crosses two operational definitions of ``bridge'' rather than one. Experiment~11 softens this, finding that degree rather than betweenness is the property that drives consequence and that the two are near-collinear in modular structure.

\begin{figure}[t]
\centering
\includegraphics[width=\textwidth]{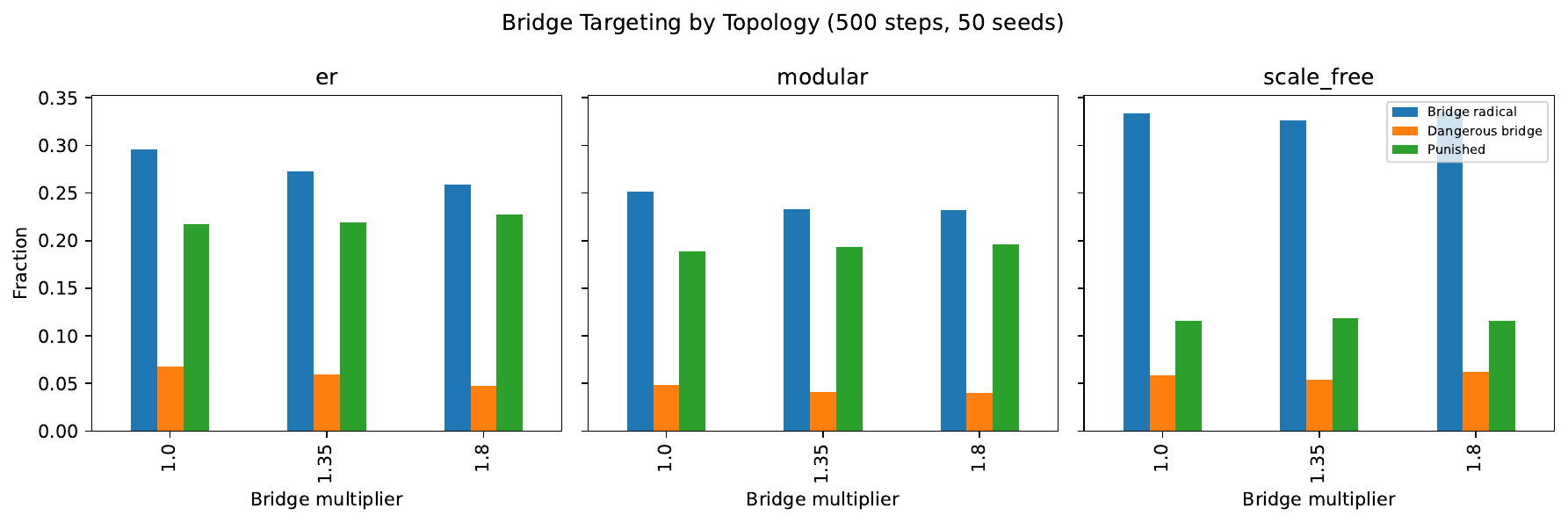}
\caption{Bridge targeting across topologies and multiplier levels ($m \in \{1.0, 1.35, 1.8\}$) at the default noise regime. Higher multipliers suppress bridge radicalization in the Erd\H{o}s--R\'{e}nyi (paired $p = 0.0008$ at $m{=}1.8$ vs.\ $m{=}1.0$) and modular ($p = 0.02$) topologies, and have no detectable effect in scale-free graphs ($p = 0.80$). \textbf{Conclusion:} targeting suppresses bridge radicalization in Erd\H{o}s--R\'{e}nyi and modular topologies but not on hub-dominated scale-free graphs (question~2); the accuracy-dependence of its net cost is examined in Experiments~4 and~9. Results averaged over 50 seeds per condition.}
\label{fig:p2_bridge_tradeoff}
\end{figure}

This experiment runs at the default noise regime; the accuracy-vs-noise contrast is examined separately in Experiments~4 and~9. The hypothesis is only partly confirmed, and not where it predicted. Bridge targeting with multiplier $m = 1.8$ suppresses bridge radicalization relative to uniform enforcement ($m = 1.0$) in two of the three topologies: the effect is largest in Erd\H{o}s--R\'{e}nyi graphs (bridge radical fraction $0.296 \to 0.259$, paired $t = 3.58$, $p = 0.0008$), present but smaller in the modular model ($0.251 \to 0.232$, paired $t = 2.40$, $p = 0.02$), and absent in scale-free graphs ($0.334 \to 0.336$, paired $p = 0.80$) (Figure~\ref{fig:p2_bridge_tradeoff}). The scale-free null comes with the definitional caveat above: its degree-defined ``bridges'' are hubs, and the design cannot separate a genuine failure of targeting from the change in bridge definition. In the modular stochastic block model, aggressive bridge targeting also reduces the dangerous-radical bridge fraction ($0.048 \to 0.040$, paired $p = 0.04$); the accompanying rise in overall punished fraction is small and not statistically significant at this sample (paired $p = 0.11$), so we treat the collateral-cost channel as suggestive rather than established here.

Whether this targeting becomes counterproductive under noisy classification (because it concentrates enforcement on exactly the nodes most likely to be misclassified) is the governance dilemma we probe in Experiments~4 and~9. There the direction is as expected (targeting trends harmful under FP-heavy noise, helpful under accurate classification), but on the governance-loss metric the effect is within noise at our sample size; we therefore state the dilemma as a directional hypothesis the data is consistent with, not a significant effect.

\subsection{Experiment~3: Governance loss decomposition}

The third experiment decomposes the governance loss $\mathcal{L}_{\mathrm{gov}}$ into its three component terms with equal weighting ($\lambda_{\mathrm{FN}} = \lambda_{\mathrm{FP}} = \lambda_u = 1$). The totals are those of Experiment~1 (Table~\ref{tab:p2_main_results}, the noise sweep at the default targeting multiplier $m = 1.35$), here split into $\mathcal{L}_{\mathrm{FN}}$, $\mathcal{L}_{\mathrm{FP}}$, and $\mathcal{L}_{\mathrm{control}}$. A separate grid that crosses the four noise regimes with four targeting multipliers ($m \in \{1.0, 1.2, 1.5, 2.0\}$; 800 runs) confirms that the oracle-to-FP-heavy ordering holds at every multiplier level (FP-heavy Cohen's $d$ between $2.1$ and $2.5$ across multipliers; the multiplier dimension itself is the subject of Experiment~4). For each of the 50 seeds per condition, the decomposition is computed from the frozen simulation outputs.

Effect sizes are computed as Cohen's $d$ between each noise condition and the oracle baseline. The oracle provides a lower bound on achievable governance loss (equal to pure control cost $\mathcal{L}_{\mathrm{control}}$ since classification is perfect). Departures from this bound quantify the governance cost attributable to classification noise, decomposed by failure mode. This experiment directly addresses the central claim that governance loss reveals costs invisible to aggregate usefulness.

\begin{figure}[t]
\centering
\includegraphics[width=\textwidth]{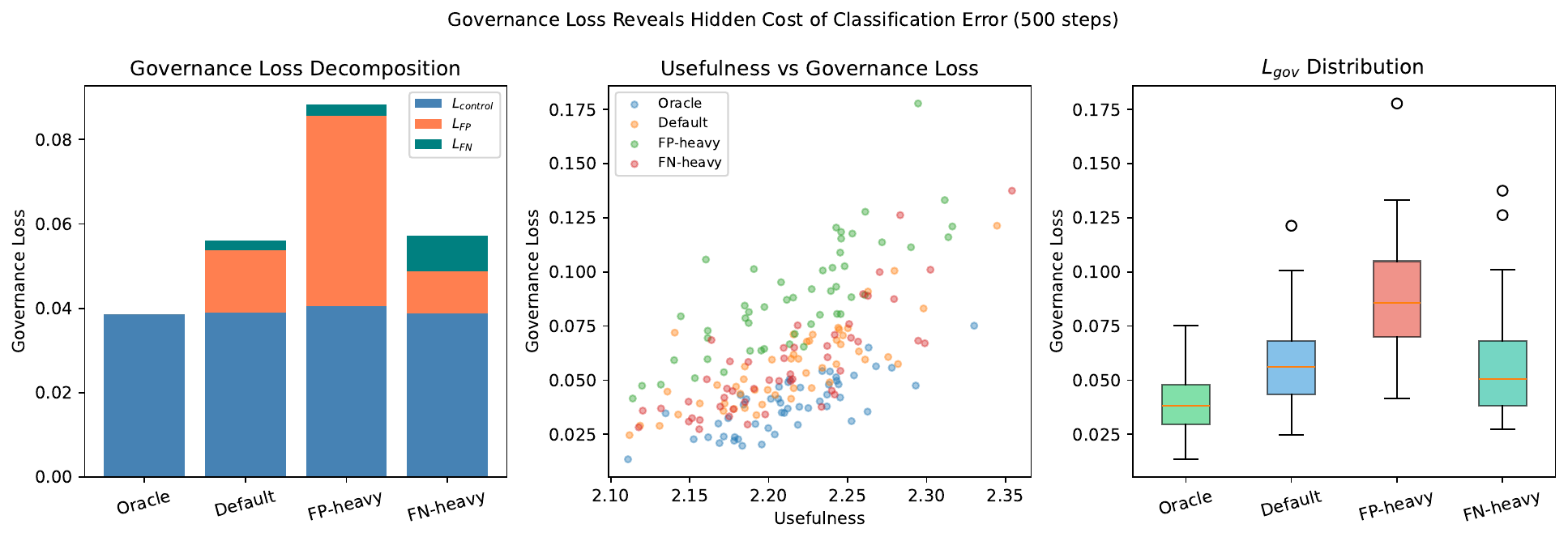}
\caption{Governance loss decomposition across noise regimes (50 seeds per condition at the default multiplier $m = 1.35$, the same runs as Table~\ref{tab:p2_main_results}). Left: $\mathcal{L}_{\mathrm{gov}}$ split into its components $\mathcal{L}_{\mathrm{control}}$, $\mathcal{L}_{\mathrm{FP}}$, and $\mathcal{L}_{\mathrm{FN}}$; FP-heavy noise inflates the false-positive bridge component ($\mathcal{L}_{\mathrm{FP}}$), FN-heavy noise the false-negative one ($\mathcal{L}_{\mathrm{FN}}$). Middle: aggregate usefulness against governance loss, with usefulness clustered near $2.2$ while governance loss spreads more than twofold. Right: the per-regime distribution of $\mathcal{L}_{\mathrm{gov}}$. \textbf{Conclusion:} governance loss separates failure modes invisible to aggregate usefulness (question~3).}
\label{fig:p2_governance_loss_decomp}
\end{figure}

The governance loss formula reveals the hidden costs that usefulness cannot detect (Figure~\ref{fig:p2_governance_loss_decomp}). Under oracle classification, governance loss is $\mathcal{L}_{\mathrm{gov}} = 0.039$, consisting entirely of control cost $\mathcal{L}_{\mathrm{control}}$ since both $\mathrm{FP}_{\mathcal{B}}$ and $\mathrm{FN}_{\mathcal{B}}$ are zero. This represents the irreducible cost of governance activity itself.

Under FP-heavy noise, governance loss rises to $\mathcal{L}_{\mathrm{gov}} = 0.088$, driven primarily by the $\mathcal{L}_{\mathrm{FP}}$ component: suppression of productive bridges. The effect size relative to the oracle is Cohen's $d = 2.41$, indicating a very large effect that would be detected with high power in any adequately sampled study. Under FN-heavy noise, governance loss is $\mathcal{L}_{\mathrm{gov}} = 0.057$ with Cohen's $d = 0.96$ versus the oracle, driven by the $\mathcal{L}_{\mathrm{FN}}$ component: undetected dangerous bridge activity. The default noise condition ($\mathcal{L}_{\mathrm{gov}} = 0.056$) falls between these extremes, with moderate contributions from both error types.

The asymmetry between FP-heavy and FN-heavy governance loss ($0.088$ vs.\ $0.057$) reflects the structure of the model: false positives on productive bridges both add to $\mathcal{L}_{\mathrm{FP}}$ directly and slightly increase effective control expenditure as the regulator acts on misclassified nodes, compounding costs in the $\mathcal{L}_{\mathrm{control}}$ term. False negatives, by contrast, represent failures to act and therefore do not inflate control costs.

\paragraph{Comparison with alternative metrics.} To assess whether $\mathcal{L}_{\mathrm{gov}}$ provides information beyond simpler alternatives, we compare eight metrics across noise conditions using one-way ANOVA (Table~\ref{tab:alt_metrics}).

\begin{table}[t]
\centering
\begin{tabular}{lccl}
\toprule
Metric & ANOVA $F$ & $p$ & Detects noise regime? \\
\midrule
Aggregate usefulness & 0.10 & 0.962 & No \\
Punished fraction & 0.17 & 0.920 & No \\
\midrule
$\mathcal{L}_{\mathrm{gov}}$ & 47.4 & $<10^{-4}$ & Yes \\
Bridge FP rate & 1843 & $<10^{-4}$ & Yes \\
Bridge FN rate & 176 & $<10^{-4}$ & Yes \\
Population FP rate & 17978 & $<10^{-4}$ & Yes \\
Population FN rate & 1317 & $<10^{-4}$ & Yes \\
\bottomrule
\end{tabular}
\caption{Metric sensitivity to noise condition (one-way ANOVA, 4 conditions $\times$ 50 seeds). Outcome-level metrics (usefulness, punished fraction) are blind to noise regime. All error-rate metrics, including $\mathcal{L}_{\mathrm{gov}}$, detect the effect. $\mathcal{L}_{\mathrm{gov}}$'s advantage is not unique detection but separation into interpretable failure modes (question~3).}
\label{tab:alt_metrics}
\end{table}

Both outcome-level metrics (usefulness and punished fraction) are blind to noise regime ($p > 0.68$). All error-rate metrics detect the effect, including population-level FP/FN rates that are not bridge-specific. The contribution of $\mathcal{L}_{\mathrm{gov}}$ is therefore not unique detection but decomposition: it combines FN exposure, FP suppression, and control cost into a single scalar whose components are separately interpretable (Desideratum~4). An analyst can read off \emph{which} failure mode dominates, not just that something differs.

\paragraph{Data provenance.} All reported values in Experiments~1--3 are computed from frozen simulation outputs generated with 50 seeds per condition on modular stochastic block model networks with bridge nodes defined by betweenness centrality. Governance loss uses equal component weights ($\lambda = 1$). Effect sizes are Cohen's $d$ computed from the cross-seed distributions.

\subsection{Experiment~4: Adaptive vs.\ static bridge targeting}

The bridge-targeting dilemma motivated by Experiment~2 (targeting may help under accurate classification but hurt under noisy classification, a contrast probed directly in Experiments~4 and~9) raises a natural question: can a regulator that adapts its targeting intensity manage this tradeoff? Experiment~4 compares four enforcement policies across four noise regimes: three static multipliers ($m \in \{1.0, 1.35, 1.8\}$) and the adaptive bandit from Section~\ref{sec:adaptive_mult}. Each combination is run for 500 steps with 50 seeds.

The hypothesis is that no single static multiplier is optimal across noise regimes, and that an adaptive bandit converges to a regime-appropriate targeting level.

The results partially confirm and partially complicate this hypothesis. Under FP-heavy noise, aggressive targeting trends worse than no targeting ($m{=}1.0$: $\mathcal{L}_{\mathrm{gov}}{=}0.087$; $m{=}1.8$: $0.090$), consistent with the governance dilemma from Experiment~2, though the gap at this single multiplier step is within noise ($t{=}0.56$, $p{=}0.58$); the clearer signal is the monotone increase across the wider $\alpha$ range in Experiment~9. Under FN-heavy noise, the ranking reverses: $m{=}1.8$ gives the lowest governance loss ($0.057$) because strong targeting compensates for missed dangerous content, though this gap is also within noise ($t{=}0.47$, $p{=}0.64$). No single static $m$ is clearly optimal across regimes.

The adaptive bandit, however, converges to $m \approx 1.50$ in \emph{all} noise regimes and does not back off under FP-heavy noise ($\mathcal{L}_{\mathrm{gov}}{=}0.091$, slightly worse than $m{=}1.0$). The reason is that its reward signal ($r_m = -\mathrm{FP}_{\mathcal{B}} - 2D_{\mathcal{B}}$) is misaligned with governance loss: bridge FP rate is determined by the confusion matrix and does not depend on $m$, so the bandit sees only that higher $m$ reduces $D_{\mathcal{B}}$ and increases its reward. It does not observe that higher $m$ also amplifies the enforcement \emph{impact} of false positives through the punishment probability, which enters $\mathcal{L}_{\mathrm{gov}}$ via the $\mathcal{L}_{\mathrm{FP}}$ and $\mathcal{L}_{\mathrm{control}}$ terms.

This is a reward-design failure, not an indictment of adaptive governance. A convergence check with 2{,}000 steps (4$\times$ the base horizon) confirms that the bandit stabilizes at $m \approx 1.5$ under FP-heavy noise: the result reflects the reward signal, not insufficient learning time. A bandit whose reward includes the enforcement cost (e.g., $r_m = -\mathrm{FP}_{\mathcal{B}} \cdot P^{\mathrm{pun}} - 2D_{\mathcal{B}} - \bar{u}^2$) would penalize high $m$ under FP-heavy noise. The finding illustrates a general challenge: the regulator must optimize governance loss, but governance loss depends on both classification quality (which the regulator cannot control) and enforcement intensity (which it can). A reward signal that conflates these two sources produces misaligned adaptation.

\begin{figure}[t]
\centering
\includegraphics[width=\textwidth]{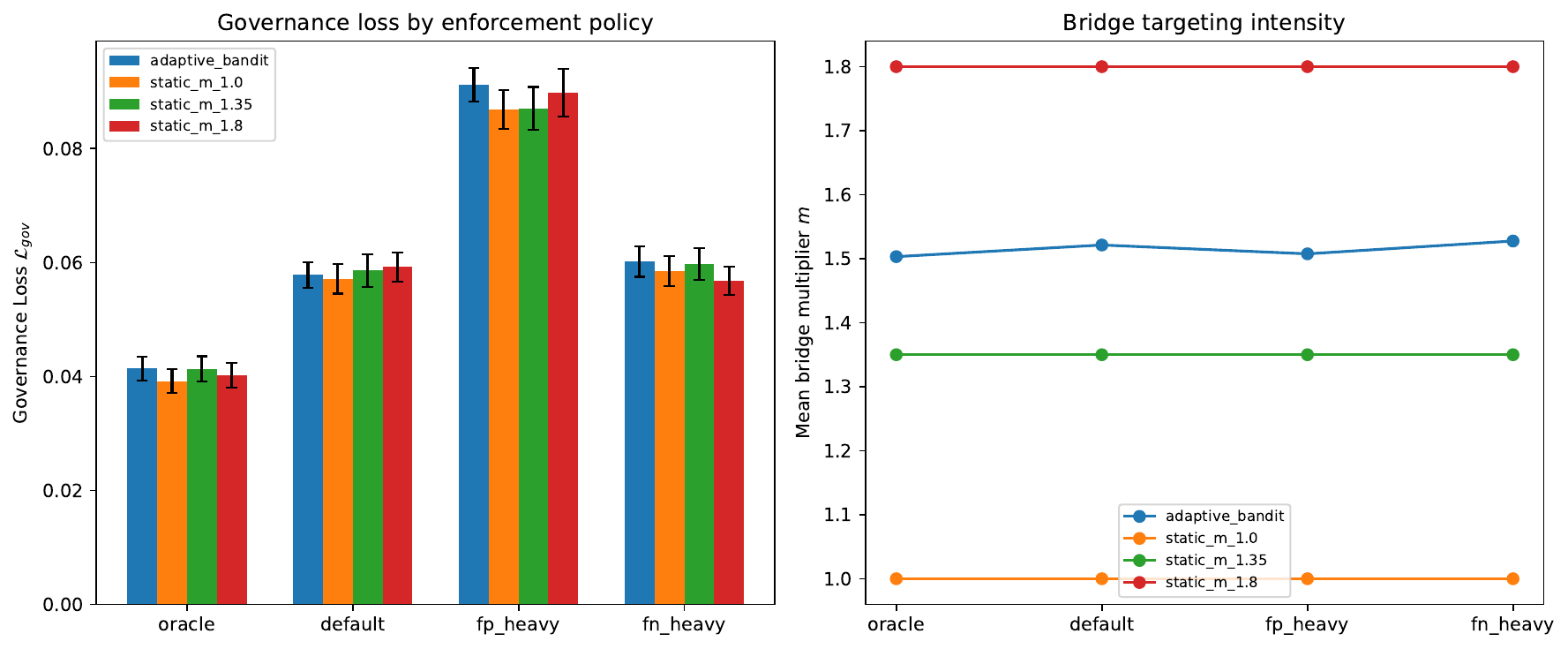}
\caption{Adaptive vs.\ static bridge targeting across noise regimes (50 seeds per condition). Left: governance loss $\mathcal{L}_{\mathrm{gov}}$ by enforcement policy and noise regime. Right: mean bridge multiplier used. \textbf{Conclusion:} the adaptive bandit converges to $m{\approx}1.5$ regardless of noise regime, failing to back off under FP-heavy noise because its reward tracks bridge error rates but not the enforcement cost of false positives (question~4).}
\label{fig:p2_adaptive}
\end{figure}

\subsection{Experiment~5: Delay $\times$ noise interaction}

The companion paper shows that institutional delay alone destabilizes otherwise stable systems. This paper shows that noise alone hides governance failures. A real regulator faces both. Experiment~5 crosses seven delay values ($\Delta \in \{0, 2, 4, 6, 8, 10, 14\}$) with four noise regimes, producing a $7 \times 4$ grid of conditions (50 seeds each, 1{,}400 runs total). The bridge multiplier is fixed at $m = 1.35$.

The hypothesis is that delay and FP-noise compound: the critical delay for instability onset is shorter under FP-heavy noise.

The results reject this hypothesis. Runaway rates are nearly identical across noise regimes at each delay: 6\% at $\Delta \leq 4$, 16\% at $\Delta{=}6$, and 30\% at $\Delta{=}8$, regardless of whether classification is oracle, default, FP-heavy, or FN-heavy. The critical delay for instability onset is set by the alarm--feedback loop (the mechanism from the companion paper) and is unaffected by classification noise.

Governance loss, however, differs sharply across noise regimes at every delay: FP-heavy noise adds ${\approx}0.05$ to $\mathcal{L}_{\mathrm{gov}}$ relative to oracle (from ${\approx}0.043$ to ${\approx}0.092$ at $m{=}1.35$), and this offset is roughly constant across all delays. Delay and noise thus operate through \emph{independent pathways}: delay drives runaway through the alarm feedback loop, while noise drives governance failure through bridge-specific classification errors. The two do not interact: they are additive, not compounding.

A two-way ANOVA confirms this independence formally: the delay~$\times$~noise interaction term is not significant for either governance loss ($F = 0.36$, $p = 0.99$, $\mathit{df} = 18, 1372$) or runaway classification ($F = 0.005$, $p = 1.00$). Delay and noise are additive. A regulator facing both faces two separate problems, not one synergistic one. Reducing delay addresses instability; improving classification addresses governance quality. Neither substitutes for the other.

\begin{figure}[t]
\centering
\includegraphics[width=\textwidth]{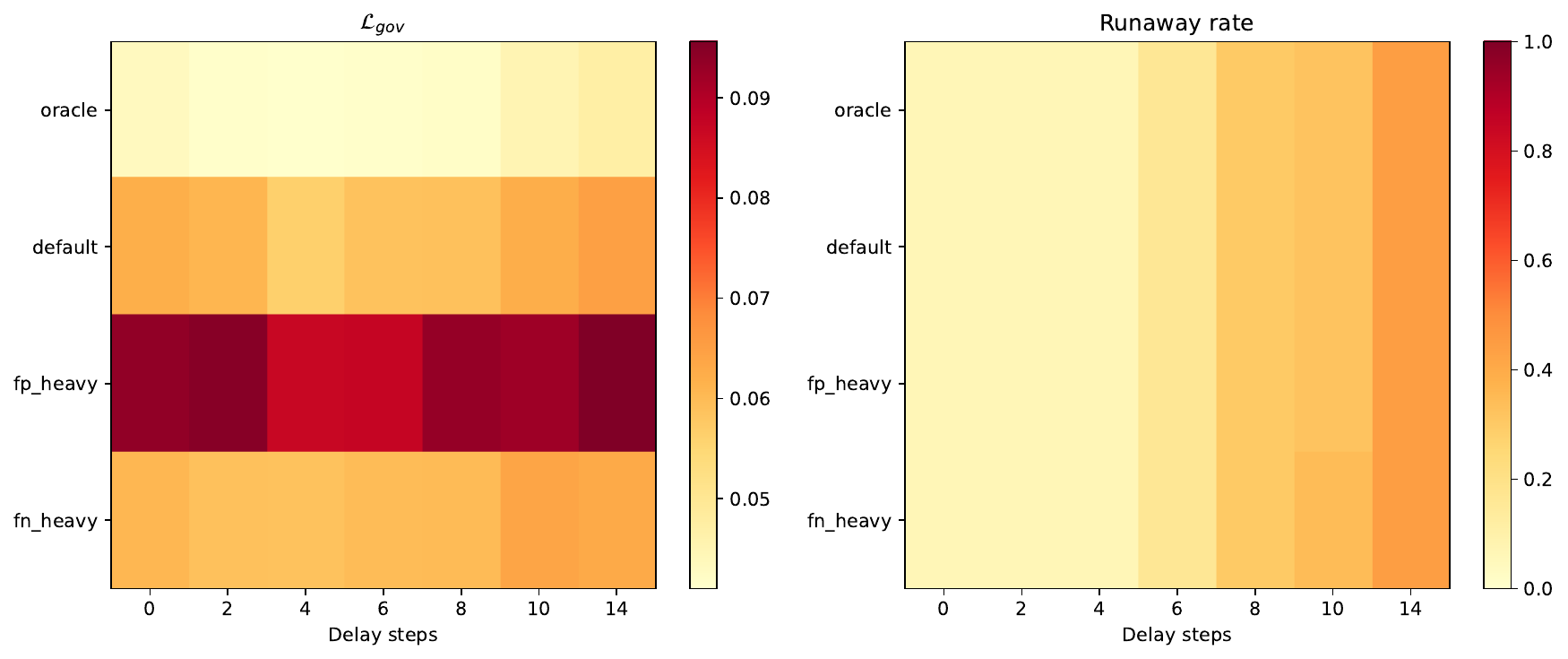}
\caption{Delay $\times$ noise interaction (50 seeds per cell). Left: governance loss $\mathcal{L}_{\mathrm{gov}}$ as a function of delay and noise regime: noise adds a constant offset at all delays. Right: fraction of runs classified as runaway. \textbf{Conclusion:} the runaway threshold is set by delay and is independent of noise regime: delay and noise act additively, not synergistically (question~5).}
\label{fig:p2_delay_noise}
\end{figure}

\subsection{Experiment~6: Endogenous content types}

In the base model, content types are fixed at initialization. Experiment~6 tests whether endogenous content dynamics change governance outcomes. Three conditions are compared: exogenous content (the base model), endogenous content with default transition strength ($\beta = 1.0$), and endogenous content with fast transitions ($\beta = 1.5$). Each is crossed with three noise regimes (oracle, default, FP-heavy), producing $3 \times 3 = 9$ conditions with 50 seeds each (450 runs).

The hypothesis is that endogenous content creates a feedback loop absent in the base model, potentially amplifying or dampening the governance failures observed in Experiments~1--3.

The main effect of endogenous content is a redistribution across types. Under the base model, the population maintains its initial composition ($H{=}0.35$, $P{=}0.45$, $D{=}0.20$). Under endogenous dynamics, the population polarizes: $H$ increases to $0.46$ ($+11$pp), $D$ increases slightly to $0.22$ ($+2$pp), and $P$ drops to $0.32$ ($-13$pp). Productive agents split: some transition toward harmless content through loyal behavior, while others who persist in radical behavior drift toward dangerous content. This polarization pattern is consistent across all noise regimes and both transition strengths, and echoes the RL-driven polarization on modular networks found by \citet{banisch2019polarization}.

Despite this population shift, governance loss under endogenous content is only marginally different from the base model: under FP-heavy noise, $\mathcal{L}_{\mathrm{gov}} = 0.082$ (endogenous, $\beta{=}1.0$) vs.\ $0.087$ (exogenous). The difference is non-significant (Welch $t = -1.04$, $p = 0.30$) and, beyond that, statistically equivalent: a two one-sided test rejects any difference larger than one pooled cross-seed standard deviation ($\varepsilon = 0.024$, $p_{\mathrm{TOST}} = 7 \times 10^{-5}$), comfortably inside the ${\approx}0.05$ oracle-to-FP-heavy gap the metric is built to register. This is a reassuring null result for the base model: the fixed-type assumption does not introduce substantial bias in the governance loss findings from Experiments~1--3. The endogenous extension changes the composition of the population---more harmless agents, fewer productive ones---but the noise--structure interaction that drives governance loss is stable under this shift.

Usefulness decreases slightly under endogenous content ($2.20$ vs.\ $2.22$) because the P-type bonus for moderate action ($+0.5$) is lost when productive agents transition to other types. We note that only moderate transition strengths ($\beta \in \{1.0, 1.5\}$) are tested; faster transitions ($\beta \geq 3$) could produce larger population shifts and different governance dynamics.

\begin{figure}[t]
\centering
\includegraphics[width=\textwidth]{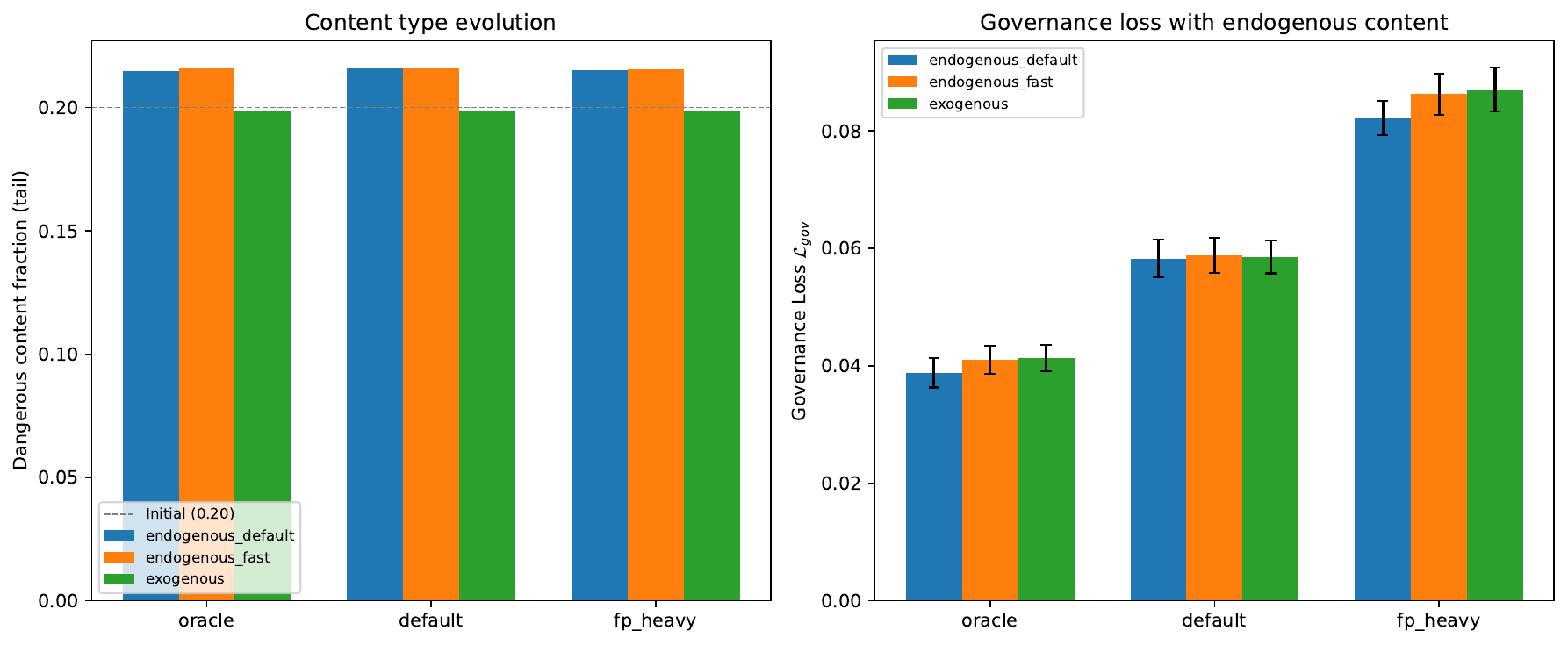}
\caption{Endogenous content dynamics across noise regimes (50 seeds per condition). Left: dangerous content fraction (D) in the tail period. The dashed line marks the initial fraction (0.20). Under endogenous dynamics the D fraction edges slightly up (to ${\approx}0.22$) and is statistically invariant across noise regimes (ANOVA $p = 0.86$ at the default transition strength; $p > 0.78$ for all endogenous subsets). Right: governance loss with and without endogenous content. \textbf{Conclusion:} endogenous content shifts the population composition but leaves governance loss nearly unchanged (question~6).}
\label{fig:p2_endogenous}
\end{figure}

\subsection{Experiment~7: Coupling strength threshold}

The usefulness invariance in Experiments~1--3 might be explained as weak coupling: usefulness computes payoffs from true content types, and classification noise affects it only indirectly through punishment $\to$ Q-learning $\to$ action shifts. If so, stronger coupling (higher punishment costs) should strengthen this indirect channel until usefulness responds to noise regime. Experiment~7 tests for such a threshold by sweeping the punishment cost multiplier over $\{0.5, 1, 2, 4, 7, 10, 15\}$ under oracle and FP-heavy noise (50 seeds each, 700 runs total).

The result is a strong null: usefulness remains statistically invariant across all seven cost multipliers ($p > 0.34$, $|d| < 0.20$ in all cases). Equivalence is confirmed directly rather than inferred from non-significance: at every cost level a two one-sided test rejects an oracle-vs-FP-heavy usefulness difference larger than $\varepsilon = 0.1$ (the Experiment~1 bound, ${\approx}2$ cross-seed standard deviations), with $p_{\mathrm{TOST}} < 10^{-12}$ throughout. Even at $15\times$ the base punishment cost, usefulness does not distinguish oracle from FP-heavy classification. Mean usefulness decreases with cost (from $2.365$ at $0.5\times$ to $1.776$ at $15\times$) because higher punishment drives agents toward low-payoff loyal behavior, but this decrease is identical under both noise regimes. The coupling threshold does not exist in the tested range: usefulness is structurally decoupled from classification noise, rather than weakly coupled.

This closes the open question within the tested range. The blindness of aggregate usefulness to classification noise is not a parameter-regime artifact that can be overcome by increasing punishment severity up to $15\times$ base cost. It is a structural property: usefulness computes payoffs from true content types, and Q-learning agents adapt their action distributions identically under oracle and noisy classification because the noise affects \emph{which} agents are punished but not \emph{how much} punishment costs. Detecting governance failures requires metrics that incorporate classification information, such as $\mathcal{L}_{\mathrm{gov}}$.

\subsection{Experiment~8: Adaptive targeting under delay}

Experiment~4 showed the adaptive bandit converges to $m{\approx}1.5$ regardless of noise regime. Experiment~5 showed delay and noise are additive. Experiment~8 tests whether the bandit remains delay-invariant: does it converge to the same multiplier under $\Delta \in \{0, 6, 14\}$ crossed with oracle and FP-heavy noise (300 runs)?

The hypothesis is that the bandit's reward signal (bridge FP rate and dangerous bridge fraction) lies in the noise-governed subspace and does not depend on delay. If so, the bandit is naturally delay-invariant without needing delay-aware modifications, a consequence of the delay--noise independence from Experiment~5.

The results confirm this: the bandit converges to $m \approx 1.50$ ($\pm 0.15$) across all six cells of the $3 \times 2$ design. Neither delay nor noise regime shifts the bandit's targeting intensity. Runaway rates increase with delay (6\% at $\Delta{=}0$, 14\% at $\Delta{=}6$, 34\% at $\Delta{=}14$) but are identical between oracle and FP-heavy at each delay, replicating the independence from Experiment~5. The adaptive bandit is delay-invariant because its reward depends on bridge-level classification outcomes (noise-governed) rather than on the alarm feedback loop (delay-governed). This result extends the delay--noise additivity from a passive observation (Experiment~5) to an active policy: a regulator adapting to noise does not need to separately adapt to delay, and vice versa. This is consistent with theoretical regret decompositions showing that delay and bandit noise contribute additively \citep{jin2022near}.

\subsection{Experiment~9: Joint optimization of targeting and classification}

Experiments~1--8 treat classifier accuracy as exogenous: the regulator adapts enforcement intensity $m$ but takes the confusion matrix $\mathbf{M}$ as given. In practice, a regulator allocates resources between two channels: improving classification (reducing FP/FN rates) and intensifying enforcement (raising $m$). This experiment asks how the optimal $m$ depends on classifier accuracy, and whether the two investments are complements or substitutes.

We parameterize classifier accuracy as a scalar $\alpha \in [0, 1]$, where the confusion matrix is the linear interpolation $\mathbf{M}(\alpha) = \alpha \mathbf{I} + (1{-}\alpha) \mathbf{M}_{\mathrm{FP}}$ between FP-heavy ($\alpha = 0$) and oracle ($\alpha = 1$). We sweep $\alpha$ from $0$ to $1$ in steps of $0.2$, crossed with $m$ from $1.0$ to $2.0$ in steps of $0.2$ (36 conditions, 20 seeds each, 720 runs). For each $(\alpha, m)$ pair, we compute the mean governance loss $\mathcal{L}_{\mathrm{gov}}$.

The dominant result is a strong main effect of classifier accuracy. Governance loss falls monotonically along the accuracy axis, at every multiplier level, from $\mathcal{L}_{\mathrm{gov}} \approx 0.089$ at $\alpha = 0$ (full FP-heavy noise) to $\approx 0.040$ at $\alpha = 1$ (oracle; minimum $0.037$ at $m = 1.6$). A regression of $\mathcal{L}_{\mathrm{gov}}$ on $\alpha$, $m$, and their interaction over all 720 runs finds the accuracy coefficient large and highly significant, the multiplier coefficient small, and the $\alpha \times m$ interaction \emph{not} significant ($\beta_{\alpha m} = -0.002$, $p = 0.73$).

The interaction we hoped to find---accuracy and targeting acting as complements (Remark~\ref{prop:complementarity})---is therefore not statistically supported. The suggestive pattern is there directionally: at $\alpha = 0$, governance loss rises with $m$ (from $0.089$ at $m{=}1.0$ to $0.093$ at $m{=}2.0$), so targeting hurts under the worst classifier; at $\alpha = 1.0$, the lowest loss occurs at $m^* = 1.6$ ($0.037$), so mild targeting helps. A corner submodularity check is consistent with complementarity ($\mathcal{L}(0, 2.0) + \mathcal{L}(1.0, 1.0) = 0.133 > 0.129 = \mathcal{L}(0, 1.0) + \mathcal{L}(1.0, 2.0)$), but the margin ($0.004$) is within cross-seed noise and 10 of 25 adjacent-corner checks go the other way. We report the accuracy main effect as the finding and the complementarity as an unconfirmed directional trend.

The practical implication rests on the main effect, not the interaction: classification accuracy is the dominant lever on governance loss, cutting it by more than half from worst to perfect classifier. Targeting intensity is a second-order knob whose sign depends weakly and non-significantly on accuracy. The cautious reading (do not target aggressively through a poor classifier) is directionally supported but not established at our sample size.

\paragraph{Limitations of Experiment~9.} With 20 seeds per cell (vs.\ 50 in Experiments~1--8), the governance loss surface is noisy: the optimal $m^*(\alpha)$ path is non-monotonic across $\alpha$, reflecting stochastic variation rather than a systematic pattern, and the $\alpha \times m$ interaction is not significant ($p = 0.73$). The robust result is the accuracy main effect; the complementarity is a directional trend that would need more seeds and a wider multiplier range to confirm.

\subsection{Experiment~10: Does bridge position amplify spread?}

The governance loss metric weights errors at bridge nodes by betweenness centrality, on the premise, imported from the influence-maximization literature \citep{kempe2003maximizing}, that bridges govern cross-community spread. This experiment tests that premise directly within our model. We place the dangerous content fraction (held fixed at 0.20) either preferentially on bridge nodes or preferentially on non-bridge nodes, under FN-heavy noise so that dangerous content is systematically missed and left free to spread influence. If bridges amplify contagion, bridge-biased placement should produce higher population-level radicalization (50 seeds per placement).

It does not. The tail-period radical fraction is $0.218$ under bridge-biased placement and $0.218$ under non-bridge-biased placement (two-sample comparison: $t = 0.12$, $p = 0.91$, Cohen's $d = 0.02$; a paired test gives the same null, $p = 0.79$). A third, uniform placement (the design also includes one; see the grid in Appendix~\ref{sec:appendix_details}) yields $0.211$, within noise of both biased placements ($p \geq 0.08$). In our model, bridge nodes' inter-community degree is too low under the sparse SBM ($p_{\mathrm{out}} = 0.004$) for bridge position to amplify population-level spread through the influence--reward coupling. This null is not an artifact of the coupling magnitude: sweeping the influence weight $\lambda_{\mathrm{infl}} \in \{0.1, 0.22, 0.5, 1.0\}$ leaves the bridge-vs-non-bridge difference negligible throughout ($|\text{diff}| < 0.003$ at all four levels: bridge $0.190/0.214/0.259/0.301$ vs.\ non-bridge $0.191/0.214/0.261/0.301$), even though stronger coupling raises overall radicalization. Nor is it an artifact of network sparsity, the parameter that most directly limits how much a bridge could amplify cross-community spread: sweeping the inter-community edge probability $p_{\mathrm{out}} \in \{0.004, 0.01, 0.02, 0.04\}$ up to ten times the base density (400 additional runs) leaves the difference within noise at every level ($|d| \leq 0.22$, all $p > 0.28$: bridge $0.218/0.225/0.238/0.268$ vs.\ non-bridge $0.218/0.224/0.241/0.267$; a TOST equivalence test bounds the difference within one cross-seed standard deviation at every density, $p_{\mathrm{TOST}} \leq 8.3 \times 10^{-5}$), even though denser inter-community links raise overall radicalization from $0.22$ to $0.27$. Bridge position confers no contagion advantage even when bridges carry substantially more inter-community traffic. The placement of dangerous content on bridges does not change population spread at any coupling we tested. This is a deliberately honest negative result: it bounds the contribution. The governance loss metric correctly detects bridge-specific errors that global metrics dilute (Experiments~1--3), but the \emph{consequence} of those errors---whether a missed dangerous bridge actually does more damage than a missed dangerous periphery node---is an assumption inherited from network theory, not a dynamic property demonstrated in this toy model. A model with explicit cascade or contagion dynamics would be required to establish that bridge errors are more costly, rather than only more visible to a bridge-weighted metric. Experiment~11 supplies that mechanism.

\subsection{Experiment~11: Does structural position carry dynamic consequence?}

Experiment~10's null has a structural cause. Influence in the base model is one-hop: an agent's reward depends only on its immediate neighbors. A one-hop quantity cannot depend on betweenness, which is defined over multi-hop shortest paths, so no amount of tuning the one-hop coupling can make bridge position matter. Experiment~11 adds the missing ingredient, the dangerous-content cascade of Section~\ref{sec:cascade}, and asks two questions: does structural position now carry outsized consequence, and is the operative property betweenness (which the metric uses) or node degree?

We place the dangerous fraction (0.20) on the highest-betweenness nodes (\emph{bridge}), on random non-bridge nodes, or on the highest-degree non-bridge nodes (a degree-matched control that holds degree high while keeping betweenness low). Each placement runs under FN-heavy noise across five contagion thresholds $\theta \in \{0, 0.2, 0.35, 0.5, 0.65\}$ spanning simple to complex contagion, at $m = 1.0$, with 50 seeds (750 runs). To give bridges enough cross-community degree for a contagion to cross, the inter-community edge probability is raised to $p_{\mathrm{out}} = 0.03$, $7.5$ times the base density. This value lies within the range Experiment~10 swept under the one-hop mechanism, but the cascade itself runs at this single density: whether position carries consequence at the base density $p_{\mathrm{out}} = 0.004$, the configuration behind the headline governance-loss numbers, remains untested, and the scope statements below carry this condition. The outcome is the peak dangerous-content fraction reached over the run, which captures reach before enforcement drives the cascade extinct; tail averages would only record the controlled steady state.

\begin{figure}[t]
\centering
\includegraphics[width=\textwidth]{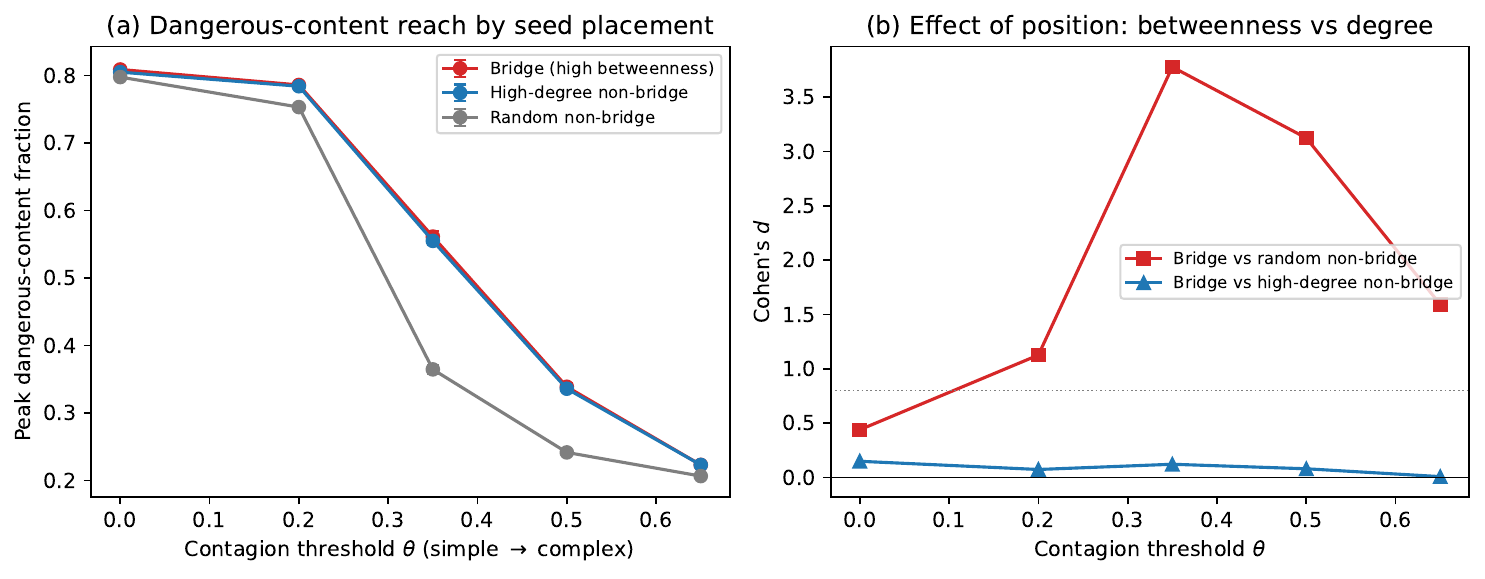}
\caption{Cascade consequence (50 seeds per cell). Left: peak dangerous-content fraction by seed placement across the simple-to-complex threshold $\theta$. Bridge and high-degree non-bridge placements track each other and both exceed random non-bridge placement, most strongly in the non-saturating (complex) regime. Right: effect size of position. Bridge versus random non-bridge is large at $\theta = 0.35$--$0.5$ (Cohen's $d > 3$); bridge versus degree-matched non-bridge is near zero throughout. \textbf{Conclusion:} structural position carries outsized consequence once content propagates, but the operative property is node degree, for which betweenness is a proxy in modular networks (question~11).}
\label{fig:p2_cascade}
\end{figure}

Structural position now matters, and strongly. Seeding dangerous content on bridges rather than on random non-bridge nodes raises peak reach from $0.365$ to $0.561$ at $\theta = 0.35$ (Welch $t = 18.9$, $p < 10^{-32}$, Cohen's $d = 3.78$) and from $0.241$ to $0.339$ at $\theta = 0.50$ ($d = 3.12$). The effect is largest in the complex-contagion regime ($\theta = 0.35$--$0.5$) and small at $\theta = 0$ ($d = 0.44$): a simple contagion saturates the population regardless of where it starts, so seed position washes out, whereas a complex contagion does not saturate and so remains sensitive to structure (Figure~\ref{fig:p2_cascade}, left). This is the scope condition for the bridge weighting: it is dynamically justified when the dangerous process does not saturate.

But the operative property is degree, not betweenness. Bridge placement and the degree-matched non-bridge control produce statistically indistinguishable reach at every threshold (Cohen's $d \leq 0.15$, all $p \geq 0.46$; e.g.\ $0.561$ vs.\ $0.555$ at $\theta = 0.35$, $t = 0.60$, $p = 0.55$; a TOST equivalence test bounds the difference within one cross-seed standard deviation at every threshold, $p_{\mathrm{TOST}} < 3 \times 10^{-5}$). Once degree is held high, betweenness adds nothing. The bridge weighting succeeds in our setting only because betweenness and degree are nearly collinear in modular networks: across seeds the two have correlation $0.96$, and bridge nodes carry $1.7$ times the degree of non-bridge nodes. Betweenness is therefore a serviceable proxy for the property that actually drives consequence (degree) in modular topologies, but the two could be separated in networks with low-degree cut vertices, where a degree-based weighting would be the correct choice.

Experiment~11 converts the imported premise of Experiment~10 into a demonstrated, scoped result. Structural-position errors do carry outsized consequence once dangerous content can propagate, provided the contagion does not saturate and the inter-community density is high enough for a cascade to cross ($p_{\mathrm{out}} = 0.03$ here); the property that makes a position consequential is its degree, which betweenness tracks closely in the modular regime the metric targets.

\section{Discussion}
\label{sec:discussion}

We read the eleven experiments as one argument about what noisy, position-blind governance does and does not change. The subsections move from the core diagnostic finding through the three model extensions to the limitations that bound it.

\subsection{Usefulness invariance as a diagnostic example}

The invariance of aggregate usefulness across noise regimes ($2.211$--$2.216$) is not a weak-coupling artifact; it is a structural property. Experiment~7 confirms this by sweeping punishment costs from $0.5\times$ to $15\times$ base: usefulness remains invariant ($p > 0.34$) at all tested levels. The mechanism has two parts, both quantifiable.

First, usefulness computes payoffs from true content types $c_i$, not predicted labels $\hat{c}_i$. Classification noise can alter usefulness only indirectly: noise changes $\hat{c}_i \to$ punishment redistribution $\to$ Q-learning update $\to$ action shift $\to$ payoff change. Each step attenuates the signal. This true-type usefulness is an idealized benchmark: it is the metric an omniscient evaluator would compute, and it is the most favorable case for the standard yardstick. A real evaluator without ground truth would estimate usefulness from the same noisy labels the regulator acts on, which would couple it to classification error directly and only widen the blind spot we document, not close it.

Second, bridge nodes constitute 12\% of the population. The additional false-positive punishment under FP-heavy noise lands on this 12\%, adding $\mathrm{FP}_{\mathcal{B}} \times f_{\mathrm{bridge}} \approx 0.23 \times 0.12 = 0.028$ to the population punishment rate, a 14\% perturbation relative to the base rate of ${\approx}0.20$. This perturbation is too small to shift the Q-learning equilibrium: agents across the population experience nearly identical punishment statistics, converge to the same action distribution, and therefore produce the same usefulness. The dilution is arithmetic, not parametric, and persists across the full coupling range tested in Experiment~7.

We should be explicit about what Experiments~1--3 demonstrate and what they do not. The governance loss metric $\mathcal{L}_{\mathrm{gov}}$ is \emph{designed} to detect bridge-specific errors, and the simulation \emph{produces} bridge-specific errors by construction (through the confusion matrix). Showing that $\mathcal{L}_{\mathrm{gov}}$ detects what it was designed to detect is a proof of concept, not an independent validation. The value of the demonstration is not that the result is surprising but that it makes the dilution arithmetic concrete and quantifiable in a complete ABM with learning agents. Experiments~10 and~11 then test whether the bridge weighting tracks a real consequence. In the base model, with only one-hop influence, it does not: placing dangerous content on bridges versus elsewhere produces no difference in population spread (Experiment~10). Adding a multi-hop contagion changes this (Experiment~11): structural position then carries large consequence in the non-saturating regime (Cohen's $d > 3$), so the weighting is dynamically justified once content can propagate. The operative property turns out to be node degree rather than betweenness, with the two near-collinear in modular networks; betweenness is a serviceable proxy here but would need to become a degree weighting where the two separate. The metric is thus a diagnostic whose weighting is demonstrated rather than assumed, under a stated scope (a non-saturating dangerous process on a modular network with inter-community density high enough for a cascade to cross).

The alternative metrics comparison (Table~\ref{tab:alt_metrics}) sharpens this point: the blindness is specific to outcome-level metrics (usefulness, punished fraction), not to all aggregate metrics. Error-rate metrics, including population-level FP/FN rates that are not bridge-specific, readily detect the effect. The contribution of $\mathcal{L}_{\mathrm{gov}}$ is therefore not detection per se, but diagnosis: it separates missed threats from suppressed coordination from control cost in a single scalar (Desideratum~4), so an analyst can identify which failure mode dominates in each noise regime.

\subsection{The governance dilemma of bridge targeting}

Bridge targeting sets a trap in principle. Under accurate classification, multiplier $m = 1.8$ does what you want: it suppresses dangerous content at structurally critical positions. Under FP-heavy classification, the same policy should backfire, concentrating enforcement on precisely those nodes most likely to be productive bridges misidentified as dangerous. Our data is consistent with this direction but does not establish it: the radicalization-suppression effect is significant in the Erd\H{o}s--R\'{e}nyi and modular topologies (though absent in scale-free, Experiment~2), while the governance-loss penalty under noise is within cross-seed noise at our sample size (Experiments~4, 9). We therefore present the trap as a mechanism the model exhibits directionally, not as a demonstrated effect.

If the dilemma holds, it implies that the value of bridge targeting is bounded above by classification accuracy at bridge positions specifically: a governance system that targets bridges aggressively should also classify them accurately. Our data is directionally consistent with this but, as noted, does not establish the noise-dependence at the governance-loss level. The intuition (you cannot aim precisely through a noisy classifier) is what Experiment~9 probes more directly.

\subsection{Adaptive governance: harder than it looks}

The adaptive multiplier bandit from Experiment~4 does not solve the governance dilemma; it illustrates why the dilemma is hard. The bandit converges to $m \approx 1.50$ regardless of noise regime, failing to back off under FP-heavy noise. The problem is reward misalignment: bridge FP rate depends on the confusion matrix, not on the multiplier, so the bandit sees only that higher $m$ reduces dangerous bridge activity. It cannot observe that higher $m$ also amplifies the enforcement \emph{cost} of those false positives.

The lesson is that adaptive governance requires a reward signal aligned with governance loss. A bandit that tracked enforcement cost alongside bridge error rates would penalize high $m$ when false positives are frequent. But constructing such a signal requires knowing the governance loss decomposition, which is the metric we proposed. The adaptive regulator and the diagnostic metric are complements, not substitutes.

More broadly, in the limit of random classification no targeting policy can outperform uniform enforcement, since the multiplier is then applied to essentially random nodes. This bound is the conceptual anchor for the accuracy main effect of Experiment~9: accuracy, not targeting intensity, is the dominant lever.

\subsection{Delay and noise are independent, not compounding}

Experiment~5 tests whether institutional delay and classification noise interact at bridge positions. They do not. \citet{miekisz2011stochastic} showed that the compound effect of delay and stochasticity in evolutionary dynamics is model-dependent; in our ABM, the two are additive.

The runaway threshold is identical across noise regimes: the fraction of seeds classified as runaway follows the same trajectory ($6\% \to 16\% \to 30\%$ at delays $0 \to 6 \to 8$) whether classification is oracle or FP-heavy. Governance loss, by contrast, differs sharply across noise regimes at every delay: FP-heavy noise adds ${\approx}0.05$ to $\mathcal{L}_{\mathrm{gov}}$ regardless of delay. The two mechanisms operate through independent pathways: delay acts through the alarm--feedback loop that determines system stability, while noise acts through bridge-specific classification errors that determine governance quality. Neither amplifies the other.

The mathematical reason is that delay drops out of the steady-state statistics. In the tail period the alarm signal is approximately stationary: $A(t) \approx A^* + \varepsilon(t)$, where $A^*$ is the equilibrium alarm and $\varepsilon$ is a stationary fluctuation. Stationarity means $A(t-\Delta)$ has the same distribution as $A(t)$, so every time-averaged functional of the delayed alarm, including the expected repression probability $\mathbb{E}[\sigma(k(A(t-\Delta) - A_c))]$, is delay-invariant; no expansion of the sigmoid is needed. Delay shifts the temporal phase of alarm fluctuations but not their distribution, so steady-state punishment statistics (and hence the bandit's reward signal) are delay-invariant. What delay does change is whether the system reaches a stationary regime at all: past the runaway threshold the alarm stops being stationary, and that is the delay pathway the companion paper analyzes.

Experiment~8 extends this from passive observation to active policy: the adaptive bandit converges to $m \approx 1.50$ at all tested delays ($\Delta \in \{0, 6, 14\}$), confirming that its reward signal lies in the noise-governed subspace. This additivity echoes regret decompositions in adversarial MDPs with delayed bandit feedback, where the delay penalty enters the regret bound as a separate additive term \citep{jin2022near, lancewicki2023delay}. The independence connects the two papers and is good news for intervention design. The companion paper shows that delay alone destabilizes an otherwise stable system; this paper shows that noise alone hides governance failures. Because the two act through separate pathways, they call for separate fixes: reducing institutional delay (the companion paper's prescription) addresses instability, improving bridge classification (this paper's) addresses governance quality, and a regulator facing both needs both.

\subsection{Classification accuracy dominates targeting intensity}

Experiment~9 sweeps classifier accuracy $\alpha$ against targeting intensity $m$. The result is asymmetric: accuracy is a first-order lever on governance loss (the $\alpha$ coefficient is large and highly significant, cutting $\mathcal{L}_{\mathrm{gov}}$ from ${\approx}0.089$ to ${\approx}0.037$), while targeting is second-order and its benefit depends on accuracy only weakly. We had expected the two to be clean complements through the product form $\mathcal{L}_{\mathrm{FP}} = \lambda_{\mathrm{FP}} \cdot \mathrm{FP}_{\mathcal{B}}(\alpha) \cdot P^{\mathrm{pun}}(m)$: raising $m$ should cost more when $\mathrm{FP}_{\mathcal{B}}$ is high, so better classification should make targeting cheaper (Remark~\ref{prop:complementarity}). But this channel requires targeting to raise the punished fraction, and in our model $P^{\mathrm{pun}}$ is nearly flat in $m$ ($\rho = 0.03$). The $\alpha \times m$ interaction is correspondingly non-significant ($p = 0.73$).

The directional pattern still points the cautious way: at $\alpha = 0$ governance loss rises with $m$ (targeting through a broken classifier hurts), and at $\alpha = 1$ a mild multiplier ($m^* = 1.6$) gives the lowest loss. The defensible policy statement is the main effect (invest in classification accuracy first) rather than a precise complementarity prescription, which our data does not establish.

\subsection{Endogenous content and feedback loops}

When content types evolve with agent behavior (Experiment~6), the system develops feedback loops absent in the base model. Enforcement shapes behavior through Q-learning, and under endogenous content this behavioral shift also changes the ground truth the classifier targets. The net compositional effect in our model is a mild \emph{polarization}: harmless content rises (to ${\approx}0.46$) and productive content falls (to ${\approx}0.32$), while the dangerous fraction edges slightly \emph{up} (from $0.20$ to ${\approx}0.22$) rather than down: some agents persisting in radical behavior drift into dangerous content faster than enforcement removes it.

This compositional shift is regime-invariant: the tail dangerous fraction does not differ across noise regimes (one-way ANOVA $F = 0.15$, $p = 0.86$). Governance loss, in turn, is barely affected: under FP-heavy noise, $\mathcal{L}_{\mathrm{gov}}$ is $0.082$ with endogenous content versus $0.087$ without, a difference an equivalence test places within one cross-seed standard deviation ($p_{\mathrm{TOST}} = 7 \times 10^{-5}$). The endogenous feedback reshapes \emph{which} content exists but not the noise--structure interaction that drives governance loss, so the fixed-type base model is an adequate approximation for the metric's behavior.

The endogenous extension connects to the opinion dynamics literature: \citet{banisch2019polarization} showed that reinforcement-learning agents on modular networks develop stable polarization through social feedback. Our model adds a governance layer: the regulator's noisy enforcement is itself a form of social feedback that shapes content evolution. \citet{salahshour2022cost} documented a similar destructive loop experimentally: noisy punishment intensifies while cooperation declines. The endogenous content model formalizes this as a feedback between enforcement, behavior, and the classification target.

\subsection{Temporal dynamics: what equilibrium averages miss}

All results in Experiments~1--3 are reported as tail-period averages over the last 250 of 500 steps. This follows standard practice in ABM reporting \citep{lee2015abm}, but it discards transient dynamics that may carry governance-relevant information.

Examining the per-step trajectories reveals that FP-heavy noise produces different \emph{temporal patterns} from oracle classification, even when the tail averages converge to similar values. Bridge FP rates under FP-heavy noise exhibit higher variance and occasional spikes where a cluster of productive bridges is simultaneously misclassified. These spikes are transient (they average out over 250 steps) but they represent governance events: periods where productive inter-community coordination is disrupted. A regulator monitoring in real time would observe these disruptions even if the time-averaged metric is within tolerance.

The adaptive bandit's multiplier trajectory provides additional temporal information, of a negative kind. The trajectory looks the same under oracle and FP-heavy classification: an early exploration phase of multiplier scatter, then convergence to $m \approx 1.5$ in both regimes (Experiment~4). The temporal record carries no noise-dependent signature, which is the temporal face of the reward-misalignment result: the bandit's reward stream contains no trace of the enforcement cost of false positives, so its trajectory cannot reflect the regime it operates in. A regulator hoping to detect the FP-heavy regime from the bandit's behavior would find nothing to read.

\subsection{Limitations}

This is a deliberately small model, and four features of it bound what the results can claim.

The governance loss carries two construction choices worth flagging. Its weights $\lambda_{\mathrm{FN}}, \lambda_{\mathrm{FP}}, \lambda_u$ are free parameters; a sweep of all 27 combinations confirms the headline conclusion is $\lambda$-robust, with only the intermediate ordering of default versus FN-heavy noise sensitive to the weighting (Appendix~\ref{sec:robustness}). The product form $\mathcal{L}_{\mathrm{FN}} = \lambda_{\mathrm{FN}} \cdot \mathrm{FN}_{\mathcal{B}} \cdot D_{\mathcal{B}}$ also sets $\mathcal{L}_{\mathrm{FN}} = 0$ whenever no dangerous content reaches bridges, regardless of the false-negative rate. This is intended (a missed threat that does not exist is costless), but it leaves the metric blind to false negatives in seeds where dangerous content happens to avoid bridges.

The content dynamics are only partly endogenous. Experiment~6 relaxes the fixed-type assumption with action-contingent transitions, yet the transition probabilities remain external parameters, identical across agents and independent of network position, community membership, or neighbor behavior. A richer model would let local social influence drive content evolution and would seed dangerous content preferentially at bridges rather than uniformly at initialization.

The base model omits cascade dynamics: a node influences its neighbors only through a weak influence--reward coupling, not through a propagating cascade, and Experiment~10 shows the consequence directly, with dangerous content on bridges spreading no further than dangerous content elsewhere. Experiment~11 adds the missing mechanism and finds that structural position then does carry outsized consequence, but three qualifications remain. First, the effect holds only when the contagion does not saturate (a simple, low-threshold contagion floods the population regardless of seed position); the bridge weighting is justified in the complex-contagion regime, not universally. Second, the property that drives consequence is node degree, which betweenness tracks closely in modular networks but not in topologies with low-degree cut vertices, where the weighting should follow degree instead. Third, the cascade runs at inter-community density $p_{\mathrm{out}} = 0.03$, raised from the base $0.004$ so that a contagion can cross communities at all; whether position carries consequence at the sparse base density, the configuration behind the headline governance-loss numbers, is untested. The cascade in Experiment~11 also keeps the transition rule external and the contagion fixed-rate rather than learned; a model in which agents choose to transmit, and in which a single missed bridge can seed a multi-community outbreak through a learned process, would tie the weighting still more tightly to the dynamics.

The structural choices are robust within the ranges we tested but untested beyond them. Repeating the core finding across bridge fractions of 5\%, 12\%, and 20\% keeps governance loss under FP-heavy noise at roughly twice the oracle value throughout (Appendix~\ref{sec:robustness}), and the usefulness invariance discussed above persists across the full coupling-strength sweep (Experiment~7). What we cannot test at $N = 240$ with idealized block structure is whether the picture holds in larger, heterogeneous networks, or where community membership is dynamic or contested and bridge positions are harder to identify. The dilution arithmetic itself, however, holds in any modular system; the numbers are model-dependent, the structural insight is not.

\section{Conclusion}
\label{sec:conclusion}

A regulator that acts on classifier labels governs a population it never observes directly. We asked what that blindness costs when classification errors fall on the few nodes that bridge otherwise separated communities, and whether the conventional yardstick (aggregate usefulness) registers the damage. It does not. Across noise regimes usefulness holds near $2.21$ (one-way ANOVA $p = 0.96$), while the bridge-weighted governance loss $\mathcal{L}_{\mathrm{gov}}$ more than doubles, from $0.039$ under oracle classification to $0.088$ under false-positive-heavy noise (Cohen's $d = 2.41$). The cause is arithmetic, not parametric: bridges are a small fraction of the population, so errors concentrated on them average away in any global statistic and survive only in a metric that weights position. The eleven experiments are best read not as a surprise but as making that dilution concrete and measurable inside a full model of learning agents.

The metric earns its place by diagnosis rather than detection. Population error rates already signal that something is wrong under noise; what $\mathcal{L}_{\mathrm{gov}}$ adds is separation, resolving missed danger at bridges, suppressed coordination at bridges, and control cost into three terms of a single scalar, so that an analyst can name which failure mode dominates in each regime.

Several of our findings are negative, and they discipline the design of noisy governance more than a run of positive results would. An adaptive bandit that targets bridges harder whenever it detects danger never learns to ease off under false-positive-heavy noise, because its reward tracks bridge error rates and not the enforcement cost those errors carry. Adaptive governance therefore needs a reward aligned with governance loss, which is the metric we propose. Institutional delay and classification noise act independently: delay sets the threshold for runaway instability, noise sets the level of governance loss, and none of our tests finds either amplifying the other. A regulator facing both must reduce delay and improve classification as separate interventions, not hope that one fix buys the other.

Two results fix the scope of the contribution. In the base model, dangerous content on bridges spreads no further than elsewhere ($p = 0.91$): with one-hop influence, bridge position confers no contagion advantage, so the weighting cannot be read off the base dynamics. Adding a multi-hop contagion changes this. Structural position then carries large consequence (Cohen's $d > 3$) whenever the dangerous process does not saturate, which demonstrates the bridge weighting rather than assuming it, under a stated scope. The property that makes a position consequential is its degree, for which betweenness is a near-collinear proxy in modular networks ($r = 0.96$) but not in topologies with low-degree cut vertices, where a degree weighting would be the right choice. The dilution arithmetic itself is generic to modular systems. Tightening the cascade into a learned transmission process, and turning the grid-search resource-allocation results into formal optimization, are the natural steps beyond this. Read this way, $\mathcal{L}_{\mathrm{gov}}$ is a diagnostic whose position weighting is demonstrated, scoped to a non-saturating dangerous process, to inter-community density high enough for a cascade to cross, and to degree as the operative axis.

\section*{Acknowledgments}
The author thanks Ilya Makarov for valuable feedback on the manuscript.

\bibliographystyle{plainnat}
\bibliography{references}

\clearpage
\appendix
\section*{Appendix}
\section{Model Parameters and Implementation}
\label{sec:appendix_details}

This appendix provides the full model specification (this section), the experimental design and statistical procedures (Appendix~B), and the robustness checks, ODD-protocol alignment, and reproducibility details (Appendix~C) that support the main text.

\subsection{Simulation configuration}

Table~\ref{tab:params} lists every model parameter and its baseline value; the rationale for the principal choices follows, and parameters varied in specific experiments are mapped in Table~\ref{tab:expgrid}.

\begin{table}[ht]
\centering
\caption{Model parameters and baseline values. Values marked $^\dagger$ are varied in the experiments mapped in Table~\ref{tab:expgrid}.}
\label{tab:params}
\small
\begin{tabular}{lll}
\toprule
Parameter & Value & Role \\
\midrule
\multicolumn{3}{l}{\emph{Network}}\\
$N$ & 240 & number of agents \\
$K$ & 6 & communities \\
$p_{\mathrm{in}}$ & 0.08 & intra-community edge probability \\
$p_{\mathrm{out}}$ & 0.004 & inter-community edge probability \\
$f_{\mathcal{B}}$ & 0.12 & bridge fraction (top betweenness) \\
\midrule
\multicolumn{3}{l}{\emph{Q-learning agents}}\\
$\alpha$ & 0.10 & learning rate \\
$\gamma$ & 0.95 & discount factor \\
$\epsilon$ & 0.08 & exploration rate \\
\midrule
\multicolumn{3}{l}{\emph{Regulator and repression}}\\
$\Delta$ & $6^\dagger$ & institutional delay (steps) \\
$k$ & 10 & sigmoid sharpness \\
$A_c$ & 0.72 & alarm threshold \\
$\lambda_{\mathrm{infl}}$ & $0.22^\dagger$ & influence coupling weight \\
$u$ & 1.0 & regulator force \\
$m$ & $1.0^\dagger$ & bridge-targeting multiplier \\
\midrule
\multicolumn{3}{l}{\emph{Actions and payoffs}}\\
$b_{L,M,R}$ & 1.0 / 2.3 / 3.0 & action benefit values \\
productive-moderate bonus & $+0.5$ & rewards the useful equilibrium \\
dangerous-radical bonus & $+0.2$ & residual payoff to radicalism \\
bridge charisma / detectability & $+0.2$ / $+0.15$ & structural-position bonuses \\
\midrule
\multicolumn{3}{l}{\emph{Content and governance loss}}\\
$(H,P,D)_0$ & 0.35 / 0.45 / 0.20 & initial content shares \\
$\lambda_{\mathrm{FN}},\lambda_{\mathrm{FP}},\lambda_u$ & $1,1,1^\dagger$ & governance-loss weights \\
$\beta$ & $1.0^\dagger$ & content-transition strength \\
\midrule
\multicolumn{3}{l}{\emph{Simulation}}\\
steps & 500 & time steps per run \\
burn-in & 250 & tail period = last 250 steps \\
seeds & $50^\dagger$ & replications per condition \\
\bottomrule
\end{tabular}
\end{table}

All experiments use a modular stochastic block model network with $N = 240$ agents partitioned into $K = 6$ communities. Intra-community edge probability is $p_{\mathrm{in}} = 0.08$ and inter-community edge probability is $p_{\mathrm{out}} = 0.004$. Bridge nodes are identified as the top 12\% of nodes by betweenness centrality computed over the undirected projection of the graph.

Agents use tabular Q-learning with learning rate $\alpha = 0.10$, discount factor $\gamma = 0.95$, and $\epsilon$-greedy exploration with $\epsilon = 0.08$. The state space encodes discretized local influence, alarm bucket, punishment status, and bridge membership. The repression parameters are: institutional delay $\Delta=6$ steps, sigmoid sharpness $k=10$, alarm threshold $A_c=0.72$, influence coupling weight $\lambda_{\mathrm{infl}}=0.22$ (a moderate value; Experiment~10 sweeps it over $[0.1, 1.0]$ and finds the bridge-contagion null insensitive to it), and static regulator force $u_t=1.0$. The action space is $\{L, M, R\}$ with benefit values $\{1.0, 2.3, 3.0\}$, chosen so that, absent punishment, radical ($R$) is the most attractive action (driving the need for governance) while the gap to moderate ($M$) is small enough that enforcement can shift the equilibrium. The content-type bonuses ($+0.5$ for productive-moderate, $+0.2$ for dangerous-radical) make productive-moderate the socially useful equilibrium; their magnitudes are illustrative and the qualitative findings are insensitive to them, since the governance-loss results depend on bridge-level classification, not on the exact payoff cardinals. Each simulation runs for 500 time steps with a burn-in period of 250 steps; all metrics are computed over the tail period. A convergence diagnostic confirms that Q-values stabilize within the burn-in: the mean radical fraction changes by less than 0.01 between the last two 100-step windows at all tested conditions (worst case: 0.009). Content types are assigned as $H$ (35\%), $P$ (45\%), and $D$ (20\%) across the population, with assignment independent of network position.

\subsection{Simulation step loop}

Algorithm~\ref{alg:step} summarizes one time step. All new features (endogenous content, adaptive multiplier) are gated by configuration flags; with them off, the loop reduces to the fixed-type base model.

\begin{algorithm}[ht]
\caption{One simulation step}
\label{alg:step}
\begin{algorithmic}[1]
\State sample predicted labels $\hat{c}_i \sim \mathbf{M}(c_i)$ for all nodes \Comment{once per node, cached}
\State read delayed alarm $A(t-\Delta)$ from the history buffer
\For{each agent $i$}
  \State $a_i \gets \epsilon\text{-greedy}(Q_i, s_i)$ \Comment{$s_i$: local influence, alarm bucket, punishment, bridge flag}
\EndFor
\If{endogenous content enabled}
  \State update $c_i$ via the action-contingent transition (strength $\beta$); resample $\hat{c}_i$
\EndIf
\State $p_i \gets \sigma\!\big(k\,(A(t-\Delta)-A_c)\big)$, scaled by $m$ if $i\in\mathcal{B}$ and $\hat{c}_i=D$ \Comment{repression probability}
\State sample punishments; compute rewards (benefit $-$ punishment $+$ influence coupling)
\For{each agent $i$}
  \State $Q_i(s_i,a_i) \gets Q_i(s_i,a_i) + \alpha\big[r_i + \gamma \max_{a'} Q_i(s_i',a') - Q_i(s_i,a_i)\big]$
\EndFor
\State append the current alarm $A(t)$ to the history buffer
\State \Return step metrics
\end{algorithmic}
\end{algorithm}

\subsection{Noise regimes}

The four noise regimes are defined by confusion matrices over content types $\{H, P, D\}$:

\begin{itemize}
    \item \textbf{Oracle:} identity matrix $\mathbf{M} = \mathbf{I}$.
    \item \textbf{Default:} moderate symmetric noise. $H \to \{H: 0.85, P: 0.13, D: 0.02\}$; $P \to \{H: 0.20, P: 0.65, D: 0.15\}$; $D \to \{H: 0.05, P: 0.20, D: 0.75\}$.
    \item \textbf{FP-heavy:} elevated $P \to D$ probability. $H \to \{H: 0.70, P: 0.10, D: 0.20\}$; $P \to \{H: 0.10, P: 0.55, D: 0.35\}$; $D \to \{H: 0.05, P: 0.20, D: 0.75\}$.
    \item \textbf{FN-heavy:} elevated $D \to \{H, P\}$ probabilities. $H \to \{H: 0.85, P: 0.13, D: 0.02\}$; $P \to \{H: 0.20, P: 0.70, D: 0.10\}$; $D \to \{H: 0.25, P: 0.50, D: 0.25\}$.
\end{itemize}

Predicted labels are sampled from the confusion matrix once per node per time step and cached for all downstream use within that step, which keeps observation, enforcement, and metric computation consistent.

\subsection{Bridge targeting}

The bridge targeting multiplier $m$ is applied to nodes in $\mathcal{B}$ whose predicted label is $D$. Three multiplier levels are tested: $m = 1.0$ (uniform enforcement), $m = 1.35$ (moderate targeting), and $m = 1.8$ (aggressive targeting). The regulator force is fixed at $u_t = 1.0$ across conditions to isolate the effect of targeting from overall enforcement level.

\subsection{Governance loss computation}

Governance loss is computed from frozen simulation outputs using the formula
\[
\mathcal{L}_{\mathrm{gov}} = \lambda_{\mathrm{FN}} \cdot \mathrm{FN}_{\mathcal{B}} \cdot D_{\mathcal{B}} + \lambda_{\mathrm{FP}} \cdot \mathrm{FP}_{\mathcal{B}} \cdot P^{\mathrm{pun}} + \lambda_u \cdot \bar{u}^2,
\]
where $D_{\mathcal{B}}$ is the dangerous-radical bridge fraction and $P^{\mathrm{pun}}$ is the overall punished fraction. All reported results use equal weights $\lambda_{\mathrm{FN}} = \lambda_{\mathrm{FP}} = \lambda_u = 1$. The product form ensures that error rates are weighted by the actual activity they fail to address.

\section{Experimental Design and Statistics}

\subsection{Experiment design grid}

Table~\ref{tab:expgrid} maps each of the eleven experiments to its varied factors, replication count, and total runs. Seeds are paired across conditions by graph realization, so differences reflect the manipulated factor rather than the network draw. The per-experiment configurations below add the detail that does not fit a table (reward formulas, transition strengths, interpolation schemes).

\begin{table}[ht]
\centering
\caption{Design grid for the eleven experiments. Total: 6{,}720 runs. Robustness sweeps reported inline (the $\lambda$, bridge-fraction, influence-weight, and $p_{\mathrm{out}}$ sweeps) are not counted here.}
\label{tab:expgrid}
\small
\begin{tabular}{cllcr}
\toprule
\# & Experiment & Varied factors & Seeds & Runs \\
\midrule
1 & Noise sweep & 4 noise regimes & 50 & 200 \\
2 & Topology $\times$ targeting & 3 topologies $\times$ 3 $m$ & 50 & 450 \\
3 & Loss decomposition & 4 noise $\times$ 4 $m$ & 50 & 800 \\
4 & Adaptive vs.\ static & 4 policies $\times$ 4 noise & 50 & 800 \\
5 & Delay $\times$ noise & 7 delays $\times$ 4 noise & 50 & 1{,}400 \\
6 & Endogenous content & 3 modes $\times$ 3 noise & 50 & 450 \\
7 & Coupling threshold & 7 punishment costs $\times$ 2 noise & 50 & 700 \\
8 & Adaptive under delay & 3 delays $\times$ 2 noise & 50 & 300 \\
9 & Joint optimization & 6 accuracies $\times$ 6 $m$ & 20 & 720 \\
10 & Bridge contagion & 3 placements & 50 & 150 \\
11 & Cascade consequence & 3 placements $\times$ 5 thresholds $\theta$ & 50 & 750 \\
\midrule
\multicolumn{4}{l}{Total} & 6{,}720 \\
\bottomrule
\end{tabular}
\end{table}

\subsection{Per-experiment configurations}

The experiments that need detail beyond the design grid are specified here; the rest use the shared settings of Table~\ref{tab:params}.

\emph{Experiment~4 (adaptive multiplier).} The adaptive bandit selects from multiplier levels $m \in \{1.0, 1.2, 1.4, 1.6, 1.8, 2.0\}$ with epsilon-greedy exploration ($\epsilon = 0.10$). Reward is $r_m = -\mathrm{FP}_{\mathcal{B}} - 2 D_{\mathcal{B}}$, updated via EMA with $\alpha_{\mathrm{EMA}} = 0.05$. The bandit operates per step, independently of the regulator's force-level selection. Four conditions are compared: three static multipliers ($m \in \{1.0, 1.35, 1.8\}$) and the adaptive bandit, crossed with four noise regimes, for 800 runs total.

\emph{Experiment~5 (delay $\times$ noise).} Seven delay values ($\Delta \in \{0, 2, 4, 6, 8, 10, 14\}$) are crossed with four noise regimes (oracle, default, FP-heavy, FN-heavy), with bridge multiplier fixed at $m = 1.35$. All other parameters match the shared settings. Total: 1{,}400 runs.

\emph{Experiment~6 (endogenous content).} Content transitions follow per-action Markov chains (Section~\ref{sec:endogenous_content}). Two strengths are tested: default ($\beta = 1.0$) and fast ($\beta = 1.5$). Three conditions (exogenous, endogenous default, endogenous fast) are crossed with three noise regimes (oracle, default, FP-heavy), for 450 runs total.

\subsection{Statistical methods}

Each condition is run with 50 random seeds (1--50). Since the same seed generates the same network graph across noise conditions, observations are paired by seed. Reported values are means across seeds; the ANOVA reported in Table~\ref{tab:alt_metrics} treats conditions as independent groups, which is conservative (a paired test would yield even larger $F$ statistics). Effect sizes are Cohen's $d$ computed as $d = (\bar{x}_{\mathrm{condition}} - \bar{x}_{\mathrm{oracle}}) / s_{\mathrm{pooled}}$, where $s_{\mathrm{pooled}}$ is the pooled standard deviation. The 50-seed design provides power $> 0.95$ for detecting effects of $d \geq 0.8$ at $\alpha = 0.05$.

\section{Robustness, ODD Protocol, and Reproducibility}

\subsection{Robustness checks}
\label{sec:robustness}

Two checks beyond the numbered experiments confirm that the core finding does not depend on specific parameter choices.

\emph{Governance-loss weighting.} The weights $\lambda_{\mathrm{FN}}, \lambda_{\mathrm{FP}}, \lambda_u$ were swept over all 27 combinations in $\{0.5, 1, 2\}$. The extremal ranking is stable in every configuration: oracle classification yields the lowest $\mathcal{L}_{\mathrm{gov}}$ and FP-heavy noise the highest. Only the intermediate ordering of default versus FN-heavy noise reverses, in 10 of 27 configurations, namely those where $\lambda_{\mathrm{FP}} \gg \lambda_{\mathrm{FN}}$ down-weights the missed-threat term. The headline conclusion that noisy classification imposes governance costs invisible to aggregate usefulness is therefore $\lambda$-robust; only the intermediate ordering needs domain-specific weight calibration.

\emph{Bridge fraction.} The bridge set is defined as the top 12\% of nodes by betweenness centrality. Repeating the core experiment with bridge fractions of 5\%, 12\%, and 20\% (50 seeds each) holds governance loss under FP-heavy noise at roughly twice the oracle value throughout: $0.082$, $0.087$, $0.100$ versus $0.035$, $0.039$, $0.048$ respectively. The qualitative finding is invariant to the bridge-definition threshold.

\subsection{ODD protocol alignment}
\label{sec:odd}

Following the ODD protocol \citep{grimm2020odd}, the model's seven elements are as follows.

\begin{description}
\item[Purpose.] Investigate how classification noise interacts with network structure to produce governance failures invisible to aggregate metrics.
\item[Entities, state variables, and scales.] $N{=}240$ agents, each with an action, a Q-table, a true content type, a predicted content type, bridge membership, charisma, detectability, and a cost scale, on a directed modular graph; one regulator holding the enforcement parameters of Table~\ref{tab:params}. One step is one interaction round; a run lasts 500 steps.
\item[Process overview and scheduling.] Each step proceeds as in Algorithm~\ref{alg:step}: (1)~noisy classification, (2)~action selection via Q-learning, (3)~optional content transition, (4)~enforcement based on the delayed alarm and predicted labels, (5)~reward computation and Q-table update.
\item[Design concepts.] \emph{Emergence}: governance failures emerge from the noise--structure interaction. \emph{Adaptation}: Q-learning. \emph{Sensing}: agents observe the delayed alarm, local influence, and punishment history. \emph{Stochasticity}: noise sampling, graph generation, and exploration.
\item[Initialization.] Graph drawn from the stochastic block model, Q-tables set to zero, content types sampled from the population distribution independently of position.
\item[Input data.] None; the model is self-contained.
\item[Submodels.] Benefit function, punishment probability, alarm computation, content-transition matrices, and the governance-loss decomposition, all defined in Sections~\ref{sec:model}--\ref{sec:governance_loss}.
\end{description}

Sensitivity analysis follows \citet{tenbroeke2016sensitivity}: the 27-configuration $\lambda$ sweep (Appendix~\ref{sec:robustness}) tests robustness to governance loss weighting, while the cross-topology experiment tests structural sensitivity. Per-step output analysis follows \citet{lee2015abm}.

\subsection{Secondary parameter magnitudes}

Several parameters enter only as secondary modifiers and are set to illustrative values: the bridge charisma ($+0.2$) and detectability ($+0.15$) bonuses (direction motivated in Section~\ref{sec:model}: structurally central nodes are both more influential and more visible), the adaptive bandit's $D_{\mathcal{B}}$ reward weight ($2\times$ the false-positive term, reflecting that missed dangerous bridges are the costlier failure) and EMA rate ($0.05$), and the confusion-matrix off-diagonals (e.g., $P{\to}D = 0.35$ under FP-heavy noise). These magnitudes are not calibrated to data; the noise regimes are comparative (oracle vs.\ default vs.\ FP/FN-heavy), and the qualitative findings depend on the bridge-level error structure rather than the exact values. The Q-learning hyperparameters ($\alpha=0.10$, $\epsilon=0.08$, $\gamma=0.95$) are standard tabular-RL defaults chosen for sufficient exploration over a moderate horizon \citep{sutton2018reinforcement}.

\subsection{Reproducibility}

All simulation code, configuration files, and outputs are stored under \texttt{experiments/}. Results reported in the main text are in \texttt{experiments/results/paper2/}. Each run produces a raw trajectory CSV, a summary with per-seed metrics, and a configuration manifest.

\end{document}